\documentclass[fleqn,usenatbib]{mnras}

\usepackage{newtxtext,newtxmath}
\usepackage[T1]{fontenc}
\usepackage{ae,aecompl}
\usepackage{rotating}

\usepackage{graphicx}
\usepackage{bm}		
\usepackage[flushleft]{threeparttable}
\usepackage{float}
\usepackage[caption = false]{subfig}
\usepackage{hyperref}
\AtBeginShipout{%
  \ifnum\value{page}>1 %
    \typeout{* Additional boxing of page `\thepage'}%
    \setbox\AtBeginShipoutBox=\hbox{\copy\AtBeginShipoutBox}%
  \fi}

\usepackage{multicol}        

\usepackage{pdflscape}
\usepackage{adjustbox}
\usepackage{threeparttable}

\title[X-ray eclipses in AGN]
      {Spectral and polarimetric signatures of X-ray eclipses in AGN}

\author[E. S. Kammoun et al.]
       {E. S. Kammoun,$^1$\thanks{E-mail: \href{mailto:ekammoun@sissa.it}{ekammoun@sissa.it}}
        F. Marin,$^2$ M. Dov\v{c}iak,$^3$ E. Nardini,$^4$ G. Risaliti,$^{5,4}$
        \newauthor and M. Sanfrutos$^{6}$\\
	$^1$SISSA, via Bonomea 265, I-34135 Trieste, Italy\\
	$^2$Universit\'{e} de Strasbourg, CNRS, Observatoire Astronomique de Strasbourg, UMR 7550, F-67000 Strasbourg, France\\
	$^3$Astronomical Institute of the Academy of Sciences, Bo\v{c}n\'{i} II 1401, CZ-14100 Prague, Czech Republic\\
	$^4$INAF - Osservatorio Astrofisico di Arcetri, Largo E. Fermi 5, I-50125 Firenze, Italy\\
	$^5$Dipartimento di Fisica e Astronomia, Universit\`{a} di Firenze, via G. Sansone 1, 50019 Sesto Fiorentino (Firenze), Italy	\\
	$^6$ Instituto de Astronom\'{i}a, Universidad Nacional Aut\'{o}noma de M\'{e}xico, A. P. 70264, 04510 CDMX, M\'{e}xico\\ }

\date{Accepted 2018 Month Day;
      Received 2018 Month Day;
      in original form 2018 Month Day}

\pagerange{\pageref{firstpage}--\pageref{lastpage}}
\pubyear{2018}

\begin{document}

\maketitle

\begin{abstract}
X-ray observations of active galactic nuclei (AGN) show variability on timescales ranging from a few hours up to a few days. Some of this variability may be associated with occultation events by clouds in the broad line region. In this work, we aim to model the spectral and polarization variability arising from X-ray obscuration events, serving as probes of the relativistic effects that dominate the emission from the innermost regions. We show that asymmetries can be clearly detected in the AGN spectra as the cloud is shading different parts of the accretion disc. We also show that these effects can be detected in the temporal evolution of the polarization degree ($P$) and the polarization position angle ($\Psi$). The variations in $P$ and $\Psi$ are highly dependent on the inclination of the system, the position of the primary source and its intrinsic polarization. Considering the disc-corona system only, for an inclination $\theta = 30$\degr\ (60\degr), $P$ increases up to $\sim 20$\% (30\%), in the 4--8~keV band, when the unpolarized primary source is obscured. However, after accounting for the contribution of parsec-scale material scattering the light in our line of sight (narrow-line region and molecular torus), the variability is smoothed out and the polarization degree can be reduced down to $\sim 1\%$ (2\%). Our results suggest that the study of eclipses in AGN with the next generation of X-ray spectral and polarimetric missions could provide unique information on the physics and structure of the innermost regions as well as of the parsec-scale material.

\end{abstract}


\label{firstpage}


\section{Introduction}
\label{sec:intro}

Active galactic nuclei (AGN) are considered to be among the best laboratories to test extreme gravity regimes. The central engine of AGN, composed of a supermassive black hole (SMBH) and its viscous accretion disc \citep{Pringle72, Shakura73}, is responsible for most of the observed, tremendous, bolometric luminosity. The X-rays are of particular interest as they are thought to be produced very close to the gravitational potential well and are imprinted with distinctive, strong signatures of special and general relativistic effects. The main X-ray spectral component of AGN spectra (known as primary emission) is likely due to Compton up-scattering of seed thermal ultraviolet (UV) photons arising from the accretion disc \citep[e.g.][]{Lightman88, Haardt93}. The primary emission spectrum is well described by a power law with a high energy cutoff. The relativistic effects are imprinted in the so-called `reflection spectrum', characterized by a prominent iron K$\alpha$ line at $\sim 6.4$\,keV (rest frame) and a strong broad excess in the emission with respect to the primary at $\sim 20-30$\,keV known as `Compton hump'. The reflection component is due to reprocessing of the primary emission by the accretion disc, which acts like a mirror, reflecting high-energy radiation rather than absorbing it \citep[e.g.][]{George91}. X-ray emission in AGN is thus expected to be polarized thanks to the scattering processes happening in the vicinity of the accretion disc, in addition to the possible polarization of the primary emission itself \citep[e.g.][]{Chandrasekhar60, Angel69, HaardteMatt93}. Absorption and line re-emission also contribute to the total polarization by diluting the signal, adding to the spectrum several local depolarized features \citep{Matt93}. The polarization signal is strongly affected by 1) the geometry of the corona-disc system \citep[e.g.][]{Schnittman09, Schnittman10} and 2) general relativistic effects that parallelly transport the polarization position angle along the photon null geodesics \citep[see e.g.][]{Matt93,Dovciak08, Dovciak11}.

Several observational evidences confirmed the existence of a relativistically blurred reflection component in AGN. In particular, X-ray spectroscopy has proven to be a powerful tool to identify the reflection features in AGN X-ray spectra, allowing us to probe the innermost regions of AGN \citep[e.g.][]{Fabian09}. Additionally, this technique may provide estimates of black hole spins \citep[e.g.][]{Risaliti13, Walton14, Marinucci14}. This was achieved thanks to the high-quality spectra provided by \textit{XMM-Newton} \citep{Jansen01} and \textit{NuSTAR} \citep{Harrison13} in the 0.3--80\,keV range. The first observational X-ray feature associated with relativistic effects was the anomalous shape of the aforementioned iron K$\alpha$ line detected in type-1 AGN, where the observer has a direct view of the central engine through the polar direction of the system \citep[e.g.][]{Pounds90,Matsuoka90}. The shape of the iron line, with an extended red wing spanning over several keV was soon associated with special and general relativistic effects blurring the signal \citep[e.g.][]{Tanaka95, Iwasawa96, Nandra97}. By fitting the observed iron line with relativistic models, it became possible to determine the spin of the black hole (BH), its mass and inclination \citep[see e.g.][for a review]{Miller07rev}. Yet, an alternative interpretation, based on partial covering absorption, has been proposed in order to explain the apparent red wing of the Fe line and the spectral curvature at hard X-rays \citep[e.g.][]{Miller08, Miller09}. The two scenarios have different advantages: on the one hand, blurred reflection is able to explain the spectral and timing properties of accreting systems for a wide range of BH mass. On the other hand, the Compton-thin to Compton-thick (and vice versa) rapid transitions that are observed in ``changing-look'' AGN \citep{Matt03} is suggestive that partial covering should be also taken into consideration. In addition, several occultation events, associated with both Compton-thin and Compton-thick clouds in the broad-line region (BLR), have been reported in AGN \citep[e.g.][]{Risaliti07, Risaliti11Mrk766, Nardini11, Sanfrutos13, Torricelli14}. {\cite{Markowitz14} estimated the probability to observe an X-ray eclipse (of any duration between 0.2~d and 16~yr) in a given source to be in in the ranges $0.003-0.166$ and $0.039-0.571$, for type I and type II AGN, respectively.} By studying X-ray eclipses, it might be possible to constrain the importance of partial obscuration, together with the geometry and location of the distant obscuring clouds. This is particularly relevant since obscuration events from BLR clouds do not affect only the AGN light curves but they show also a strong impact on their spectroscopic and polarimetric properties. \cite{Risaliti11} investigated the effects of successive eclipses of the receding and approaching parts of the accretion disc on the shape of the iron line. In fact, due to special and general relativistic effects, obscuring various parts of the disc will result in a variability in the profile of the observed emission line, which provides a new probe of the innermost regions of the disc. In a similar fashion, \cite{Marin15} explored the effects of such events on the polarimetric signal. The authors showed that eclipses induce a variability in the polarization signal due to the covering of different parts of the disc emitting a non-uniformly polarized light, mainly due to relativistic effects. 

\begin{figure}
\centering
\includegraphics[width = 0.95\linewidth]{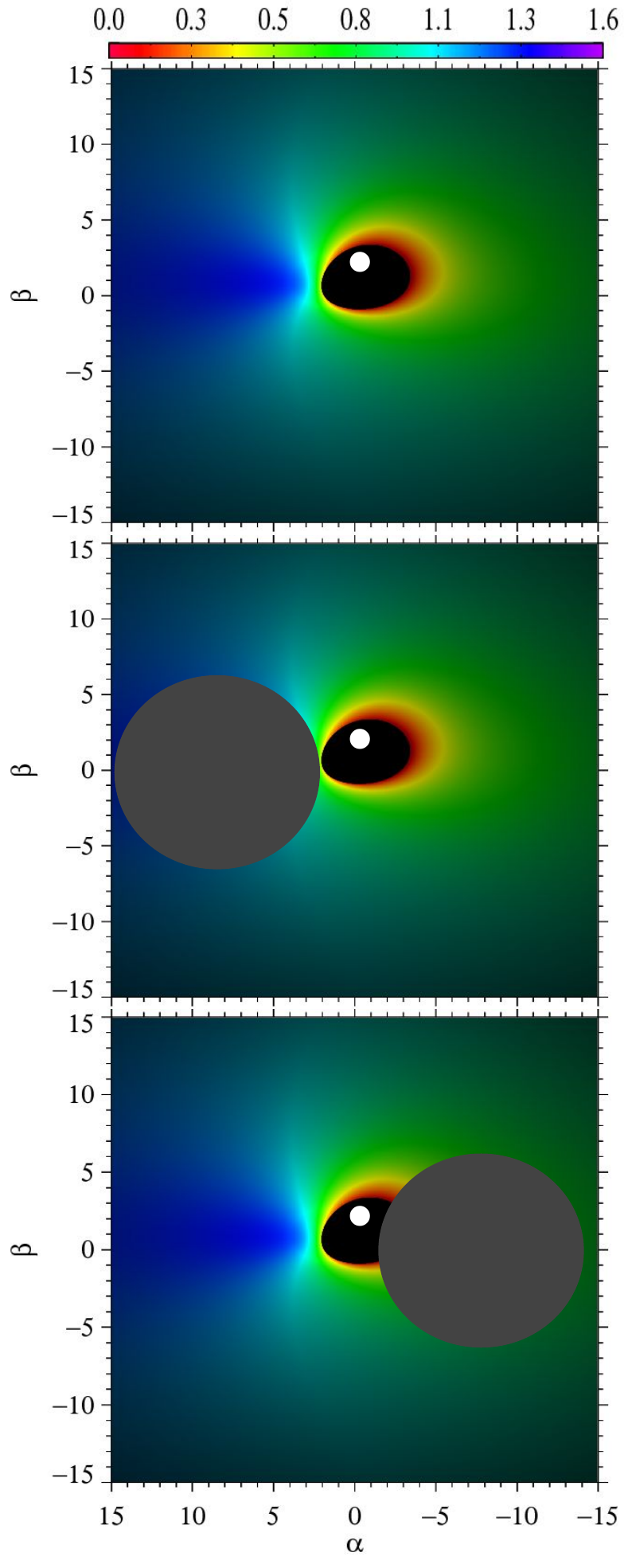}
\caption{Scheme of the configuration which we propose in this work, depicting a spherical cloud moving along the line of sight and shading different parts of the accretion-disc and lamp-post (white dot) system. The colour scale represents the energy shift defined as the ratio observed energy over local energy ($g =E_{\rm O}/E_{\rm L}$). $\alpha$ and $\beta$ represent the coordinates on the sky of the observer in units of $\rm r_g$ (impact parameters) in the $\varphi-$, $\theta-$ direction, respectively.
}
\label{fig:sketch}
\end{figure}

These pioneering studies predict what a time-resolved spectroscopic or polarimetric observation would detect, provided that the source is bright enough and that a random occultation event is serendipitously caught. Both the aforementioned papers considered a simple geometry for the accretion disc, divided in a receding and an approaching halves, illuminated by a power-law primary continuum. It is now possible to refine those studies by applying the state-of-the-art relativistic modelling to account for a more complex disc flux pattern, together with a better constrained coronal emission. In this work, we build upon the previous analyses by \cite{Risaliti11} and \cite{Marin15} in order to estimate the spectral variability as well as the time-dependent polarization that are induced by obscuration events. In this paper, we use a full relativistic ray-tracing model that allows us to track the position of a spherical cloud as it obscures different regions of the disc. The disc is illuminated by a point-like source located on the axis of rotation of the BH, known as lamp-post geometry. We describe our model in Section \ref{sec:model} and present in Section\,\ref{sec:spectroscopy} the spectroscopic signatures of various eclipsing events. We present in Section \ref{sec:Polarization} the variability pattern induced by these events in the polarization signal. We show also, in the same section, how the polarization signal may be altered by accounting for a more complex and realistic AGN model, namely by adding the contribution to polarization from the narrow-line region (NLR) and a molecular circumnuclear torus. We discuss the implications of our results in terms of observability in Section~\ref{sec:discussion} and present our conclusions in Section\,\ref{sec:conclusions}.

\begin{figure*}
\centering

\includegraphics[width = 0.95\linewidth]{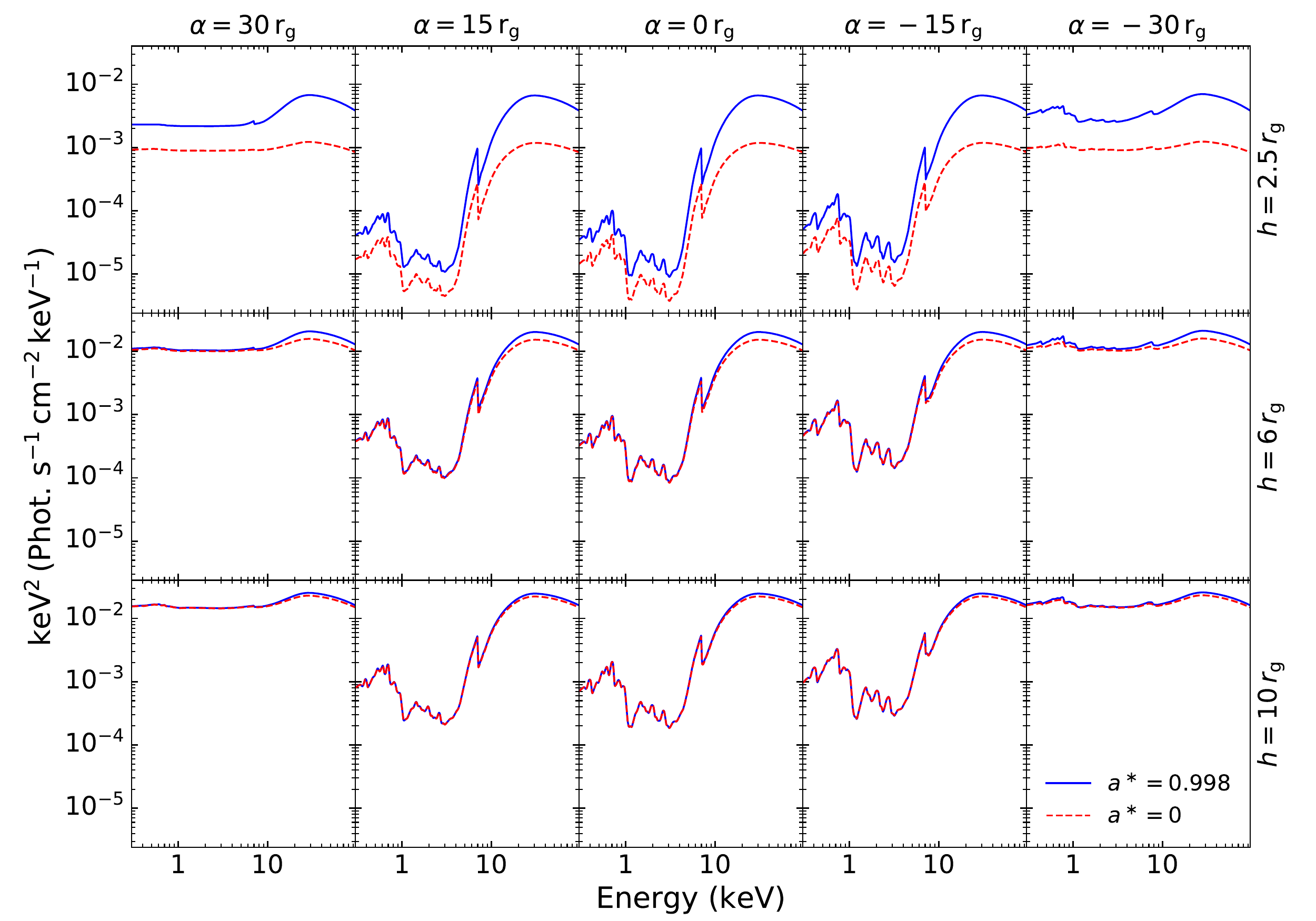}
\caption{The observed spectra resulting from obscuration 
	  events as a clouds of radius $\rm 30\, r_g$, located
	  at a typical BLR radial distance and orbiting Schwarzschild 
	  ($a^\ast = 0$; dashed red lines) and Kerr ($a^\ast = 0.998$; 
	  solid blue lines) BHs, is passing through the observer's
	  line of sight (i.e. for different values of $\alpha$, assuming $\beta =0$). We 
	  consider primary sources located at 2.5, 6 and 10 $\rm r_g$ above an accretion disc with
	  an inclination of 60\degr.}
\label{fig:spectra}
\end{figure*}
\begin{figure}
\centering

\includegraphics[width = 0.99\linewidth]{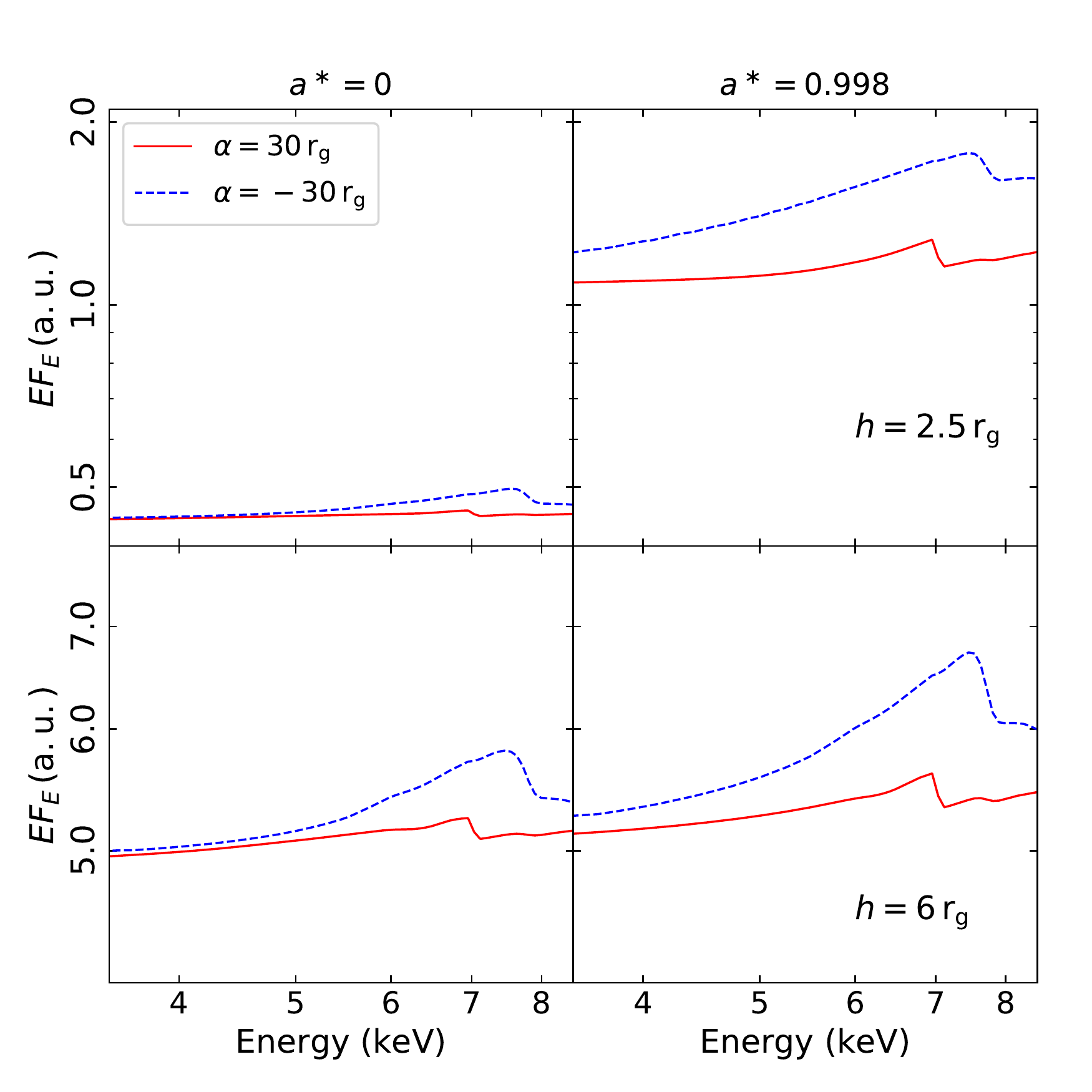}
\caption{The observed spectra, in the 3.5-8.5~keV range, considering 
		a cloud of radius $\rm 30\, r_g$ obscuring the approaching 
		($\alpha = 30\,\rm r_g$, solid red lines) and receding 
		($\alpha = -30\,\rm r_g$, dashed blue lines) parts 
		of the accretion disc for BH spins of 0 (left panels) and 0.998 (right panels). 
		We consider primary sources located at 2.5 and 6 $\rm r_g$ (top and bottom 
		panels, respectively)
		above an accretion disc with an inclination of 60\degr.}
\label{fig:FeKspectra}
\end{figure}
\section{Model}
\label{sec:model}

Our model consists of a central BH of mass $M_{\rm BH}=10^7\,\rm M_\odot$ that is accreting matter through an optically thick geometrically thin disc. This mass is typical for nearby Seyfert galaxies \citep[see e.g.][]{Lubinski16}. The disc is inclined by $\theta =30$\degr\ or 60\degr\ with respect to our line of sight.
We consider a Schwarzschild BH (i.e., with a dimensionless spin parameter of 0) and a maximally rotating Kerr BH (with spin 0.998). The primary source is assumed to be point-like (lamp-post), located on the rotation axis of the BH, emitting a power-law spectrum with a photon index $\Gamma = 2$. We investigate in this work the effects of the height of the source by considering a source at heights 2.5, 6 and 10 r$_{\rm g}$ ($\rm r_g = GM/c^2$). The higher the height the weaker the gravitational redshift and the light bending effect, which will lead to illuminating further out regions of the disc by the lamp-post. The X-ray spectra are simulated using different flavours of the relativistic ray-tracing KYN\footnote{\url{https://projects.asu.cas.cz/stronggravity/kyn}} code \citep{Dovciak14}, which computes the time evolution of a local spectrum seen by a distant observer. The model allows us to consider obscuration by a spherical cloud, as it crosses the line-of-sight of the observer. The cloud is characterized by two numbers $(\alpha,\,\beta)$, which represent the position of the centre of the cloud on the sky of the observer (impact parameters of the centre) in the azimuth and altitude ($\varphi$, $\theta$) direction, respectively, and by its radius $R_{\rm c}$. We assume a cloud of radius $30\,\rm r_g$ that is moving with a Keplerian velocity ($v_{\rm K}$), in the plane $\beta = 0$, and located at a distance of $10^4\,\rm r_g$ from the central BH \citep[see e.g.][]{Risaliti09, Marinucci14} moving from $\alpha >0$ (the approaching part of the disc) toward $\alpha <0$ (the receeding part of the disc). This will result in a duration of the eclipse $\Delta t = 2R_{\rm c}/v_{\rm K} \simeq 295$~ks. This is in agreement with the duration of the eclipse in MCG--06-30-15 \citep[$M_{\rm BH} \simeq 1.6 \times 10^6\, \rm M_\odot$;][]{Bentz16}, which was reported by \cite{McKernan98}, when scaled for the mass of the source. A scheme of the corona-disc-cloud system as projected in the observer's sky is presented in Fig. \ref{fig:sketch}. 

\begin{figure}
\centering

\includegraphics[width = 0.99\linewidth]{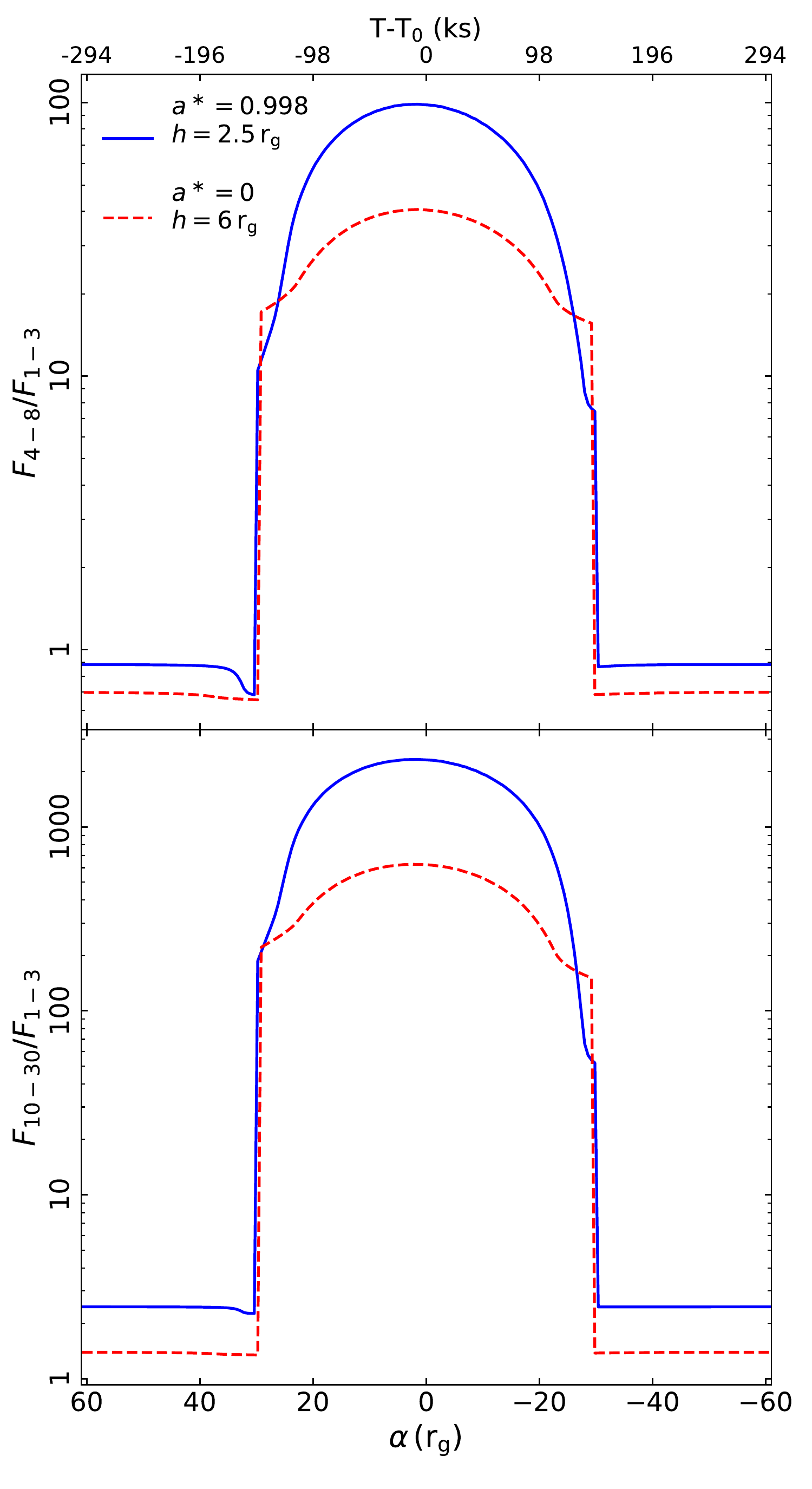}
\caption{Examples of the hard-to-soft flux ratios expected from X-ray eclipses by a cloud 
		of radiux  $\rm 30\, r_g$, as it moves along the line of sight. We consider the soft 
		band in the 1--3~keV range as a proxy of the primary emission, and two hard bands 
		the 4--8 (top panel) and 10--30~keV (bottom panel) bands, which should be dominated 
		by the FeK line and the Compton hump, respectively. We show the cases of a maximally 
		rotating BH with a lamp-post located at 2.5 $\rm r_g$ (solid blue lines) and a 
		Schwarzschild BH with a lamp-post located at 6 $\rm r_g$ (dashed red lines).}
\label{fig:HR}
\end{figure}


\section{Spectral signatures}
\label{sec:spectroscopy}

In this section, we compute the predicted spectral signatures of an X-ray eclipse by a Compton-thick cloud orbiting the BH. We assume the following model:
$${\tt model = TBabs \times KYNXillver_{abs.} + KYNXillver_{non-abs.} },$$
\noindent
where {\tt TBabs} \citep{Wilms00} represents a spherical cloud of column density $N_{\rm H} = 10^{24}\rm cm^{-2}$  (equivalent, for $R_{\rm C}=30~\rm r_g$ and $M_{\rm BH} = 10^7~\rm M_\odot$, to a number density $\sim 10^{10}\,\rm cm^{-3}$, which is consistent with the expected values for the BLR clouds), while ${\tt KYNXillver_{abs.} }$ and ${\tt KYNXillver_{non-abs.} }$ represent the spectrum that is transmitted\footnote{We note that we considered the photoelectric absorption only, neglecting Compton scattering out of and into the line of sight.} through the cloud and the non-absorbed spectrum, respectively. The reprocessing of the primary emission in the ionised disc is estimated according to the {\tt \small XILLVERD} tables \citep[see][]{Garcia16}. We considered an ionized disc with density $n_{\rm H} = 10^{15}\,\rm cm^{-3}$, ionization parameter $\log \xi_{\rm d} = 2$, and solar abundance, observed at inclinations of $\theta = 30$\degr\ and $60$\degr. Fig.\,\ref{fig:spectra} shows the resulting spectra for different positions of a cloud of radius $\rm 30\, r_g$ as it passes in the line of sight for Schwarzschild and Kerr BHs (dashed red lines and solid blue lines, respectively) and lamp-post heights of 2.5, 6 and 10\,$\rm r_g$, for $\theta = 60$\degr. The spectral variability for the case of $\theta = 30$\degr\ shows a very similar trend compared to the 60\degr\ case.  

A clear decrease in the flux below $\sim 20$~keV can be observed when the lamp-post is obscured. The source becomes reflection dominated, since the main contribution is from the unobscured parts of the disc. Below 4~keV, the flux decreases on average by $\sim 30$ and 100 times, for $h=10\,\rm r_g$ and $h=2.5\,\rm r_g$, respectively. In the Fe line band (5-7 keV) we see a smaller decrease by a factor of $\sim 7-8$, which is similar for all heights. By comparing positive and negative values of $\alpha$, this figure clearly shows the asymmetry that is introduced by the relativistic effects as the cloud obscures different parts of the disc. When the cloud is obscuring the receding patches of the disc ($\alpha < 0$) the intensity is typically higher compared to its symmetric position when the cloud is obscuring the approaching patches of the disc ($\alpha > 0$).

We note that the differences in spectral shape and intensity between a BH of spin 0 and 0.998 are larger for lower heights. It is clear from the upper row of Fig.~\ref{fig:spectra} that the spectra corresponding to the Kerr BH are brighter than the ones corresponding to a Schwarzschild BH. The difference in intensity for $h=2.5~\rm r_g$ is mainly due to the fact that, in the latter (Schwarzschild) case, the primary is much more diluted due to the source being very close to the event horizon (only distant 0.5~$\rm r_g$ as opposed to 1.5~$\rm r_g$ in the former, Kerr, case). Moreover, there is less reflection due to the fact that the innermost stable circular orbit is $\sim 5$ times farther from the BH. These differences become less prominent for larger heights as light bending gets weaker and the lamp-post can illuminate further out regions of the accretion disc. This is in agreement with previous spectral analyses \citep[e.g.][]{Fabian14, Dovciak14, Kammoun18}. 

We note also that the profiles of the emission lines differ when different parts of the accretion disc are shaded, as shown in  Fig.\,\ref{fig:FeKspectra}. This figure presents a zoom-in on the observed spectra in the 3.5--8.5~keV band, which is dominated by the Fe K line, for the cases when the cloud is obscuring the approaching ($\alpha = +30\,\rm r_g$) and receding ($\alpha = -30\,\rm r_g$) parts of the disc. We considered the cases of Schwarzschild and Kerr BHs, for lamp-posts at $h=2.5$ and $6\,\rm r_g$ above the accretion disc with an inclination of 60\degr. In addition to the effects mentioned above, this figure clearly shows the variability in the profile of the Fe line owing to shading various regions of the accretion disc. When the cloud is obscuring the receding parts of the disc, the unobscured (approaching) parts of the disc will dominate the observed spectrum. In this case, the spectrum is brighter and the emission line is blueshifted with respect to the case when the cloud is obscuring the approaching part of the disc, thanks to Doppler boosting and Doppler shift effects.

Moreover, our model allows us to estimate the time evolution of the hard-to-soft ratio (HR) light curves during eclipses. Fig.\,\ref{fig:HR} shows examples of the HR light curves of the fluxes in the 4--8 keV band (characterized by the Fe K line) and the 10--30~keV band (characterized by the Compton hump) over the one in the 1--3~keV band (representing the primary emission) during the eclipse by a cloud of radius 30~$\rm r_g$. We consider the cases of a maximally rotating Kerr BH with a lamp-post at 2.5~$\rm r_g$ and a Schwarzschild BH with a lamp-post at 6~$\rm r_g$ above an accretion disc with an inclination of 60\degr. Considering the first case, the HRs are constant when the cloud is obscuring the outer parts of the disc. A small dip can be then observed for $\alpha \simeq 32~\rm r_g$ as the cloud obscures the innermost regions of the disc, where Doppler boosting is maximum, leading to a decrease of the flux in the hard bands, where reflection from the disc is important. This dip is smaller for the 10--30~keV band, which will be affected less by absorption for the considered column density ($N_{\rm H} = 10^{24}\,\rm cm^{-2}$). Once the primary source is obscured, the HRs increase suddenly due to the decrease of the flux in the 1--3~keV band. As the cloud moves into the line of sight, the HR increases until reaching $\alpha \simeq 1.5~\rm r_g$ then starts decreasing again. We note that the HR plots show an asymmetric profile, due to relativistic effects as the cloud obscures different regions of the accretion disc. A qualitatively similar behaviour can be also observed for the case of $a^\ast = 0$ and $h=6\rm ~ r_g$. The main differences compared to the previous case are that 1) the obscuration of the primary occurs later due to the fact that the lamp-post is at a larger height and 2) the dip at $\alpha \simeq 32~\rm r_g$ is less prominent due to the fact that for low spins the disc does not extend close to the BH, thus the innermost regions which are supposed to be affected the most by relativistic effects do not contribute to the observed signal. HR light curves were also studied by \cite{Sanfrutos16}, who assumed instead a radially extended corona. {We note that many AGN, at low redshift, show large amplitude variability on timescales of a few tens of kiloseconds. MCG--6-30-15, for example, showed an intensity change by a factor of $\sim 4$ within $\sim 10$~ks \citep[e.g.][]{Fabian02}. When this variability is caused by a change in the intrinsic luminosity of the primary with a constant power-law slope, the HR light curves are expected to be constant. However, some sources may also reveal a variability in the power-law slope on timescales of a few tens of kiloseconds, as it is the case for IRAS 13224--3809, which shows a positive correlation between the power-law slope and the brightness of the source, i.e. the brighter the source, the softer the power-law \citep[e.g.][]{Kammoun15, Jiang18}. A similar behaviour was observed on longer timescales \citep[$\sim 7-11$~yr; e.g.][]{Sobolewska09}. For such variability scenario, the HR will be larger when the source is dimmer, showing a similar behaviour to the one expected from X-ray eclipses. However, the shape of the HR strongly depends on the variability pattern and such analysis is beyond the scope of this paper.}

\section{Polarimetric signatures}
\label{sec:Polarization}

\begin{figure}
\includegraphics[width = \linewidth]{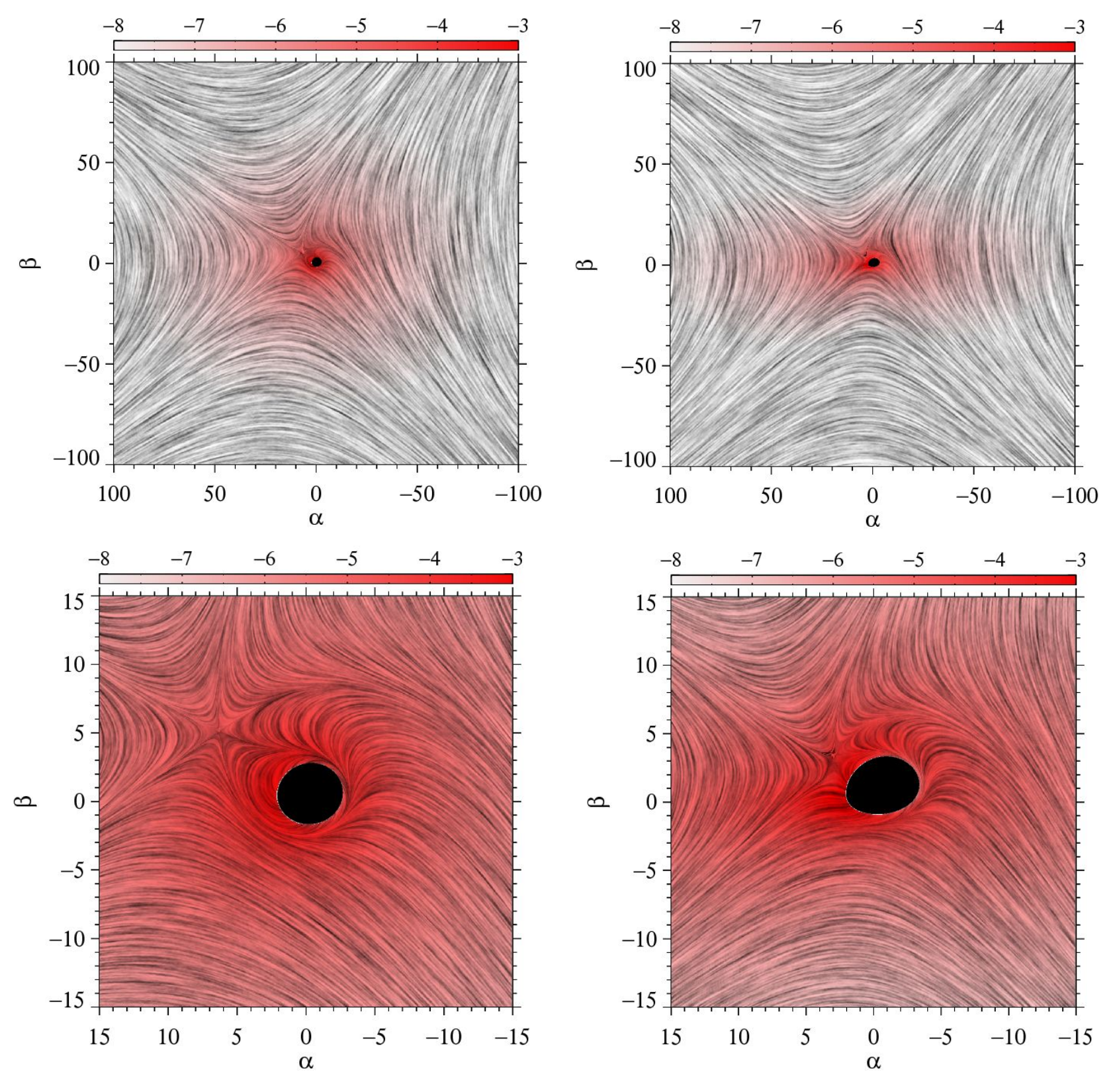}
\caption{The expected polarization, in the 2--8~keV range, throughout the accretion disc (projected in the observer's sky) for a lamp-post located at 2.5~$\rm r_g$ above a Kerr BH. We consider inclinations of 30\degr\ (left panels) and 60\degr\ (right panels). The black curves depict the direction of the polarization at the detector. It is determined by the local polarization direction induced by disc reflection (for an unpolarized primary) as well as the rotation of the polarization vector between the disc and observer due to relativistic effects. The colour scale represents the polarized flux (defined as total flux times polarization degree, in log scale). The lower panels show a zoomed in snapshot of the innermost regions of the disc, where the relativistic effects dominate.}
\label{fig:polflux}
\end{figure}

In this section, we present the effects of obscuration events on the polarization signal. For this reason we use the {\tt KYNlpcr} model that allows us to obtain the Stokes parameters ($I, Q, U, V$), the polarization degree ($P$) and the polarization position angle ($\Psi$) as a function of energy (in the 2--60 keV range), for each position of the cloud. The polarization degree is computed according to the usual Stokes formalism, i.e., 
$$P = \frac{\sqrt{Q^2+U^2+V^2}}{I}$$ 
\noindent and the polarization position angle is 
$$\Psi = \frac{1}{2}\arctan\left(\frac{U}{Q}\right).$$ 

\noindent We define the polarization position angle with respect to the system rotation axis, so a value of 0\degr\ (90\degr) will be referred to as parallel (perpendicular) polarization. For this case, the model assumes a neutral accretion disc based on the {\tt NOAR} tables \citep{Dumont00}, for spectral features of the disc reflection, with single scattering approximation \citep{Chandrasekhar60} for local polarimetric properties. We note that reflection tables for the polarization of ionized discs are not available in the literature yet (Goosmann, Marin et al., in prep). {The effect of disc ionization would be imprinted mainly in the unpolarized emission lines characterizing the reflection component, affecting mainly the soft X-rays.} The cloud is assumed to be optically thick ($N_{\rm H} \rightarrow \infty$). In reality, the contribution of the transmitted light, through the cloud, to the polarization signal is negligible \citep[see][]{Marin15}, hence it can be safely neglected. We consider a primary source at heights 2.5, 6 and 10 $\rm r_g$, being either unpolarized or polarized with $(P, \Psi)$ = [(2\%, 0\degr), (2\%, 90\degr)]. The real polarization state of the primary source depends on its geometry, properties (temperature and optical depth), and the viewing angle. Thus we chose these states as an intermediately polarized light, which is expected from a symmetric corona \citep[see][for more details]{Schnittman10}. We considered both Schwarzschild and Kerr BHs and inclinations of 30\degr\ and 60\degr. 

Fig.~\ref{fig:polflux} shows two examples of the pattern of the polarized light throughout an accretion disc for the two considered inclinations, assuming a lamp-post at 2.5~$\rm r_g$ above a Kerr BH. The zoomed out snapshots (upper panels) show that the contribution to the polarized signal is mainly arising from the innermost regions of the accretion disc. Moreover, it is clear from these panels that, due to the asymmetry caused by the non-zero inclination, the contributions from the `equatorial' regions of the disc are larger than the ones from the `polar' zones. We note that this effect is higher for larger inclinations. This figure also shows the dependence of the polarization angle on the location of the emission from the disc. The polarization induced by the scattering process has a perpendicular direction to the scattering plane. Thus the polarization of the rays emitted from the on-axis primary source and scattered from the accretion disc at large radii, where both the special (due to relatively slower orbital speeds) and general relativistic effects (due to the smaller gravity) are weak, will depend only on the azimuth, which determines the scattering plane. For example, the light rays emitted towards or away from the observer that scatter from the accretion disc and reach the observer (polar regions of the accretion disc in Fig.\,\ref{fig:polflux}, close to $\alpha=0$) have a vertical plane of scattering, and thus the polarization direction is horizontal. On the other hand, the light rays emitted to the east or west side of the accretion disc that scatter towards the observer (equatorial regions of the accretion disc in Fig. \,\ref{fig:polflux}, close to $\beta=0$) have a horizontal scattering plane, and thus the polarization direction is vertical. 

We also note the presence of a depolarizing region \citep[a.k.a. ``critical point", see][]{Dovciak08} located to the North-West of the BH ($\alpha >0$ and $\beta >0$ in Fig.\,\ref{fig:polflux}), where the photons are emitted perpendicularly to the disc, due to special relativistic aberration. The location of this region depends on the spin of the BH and on the inclination angle, and it may get closer, or even within the ISCO in some cases \citep[see Fig.\,3 in][]{Dovciak08}. Moving closer to the BH from the critical point, the relativistic effects are strong enough to affect the incident and reflection angle of the photons in the local co-moving frame. Therefore also the orientation of the local scattering plane and consequently of the polarization direction will be altered with respect to the cases where relativistic effects are not as prominent. The polarization direction is further changed as the polarization vector is transferred from the inner accretion disc to the observer at infinity. Hence the observed direction of polarization from these regions is not trivial (see the central parts of the accretion disc in Fig. \,\ref{fig:polflux}).
 

\begin{figure*}
\centering
\includegraphics[width = 0.329\linewidth]{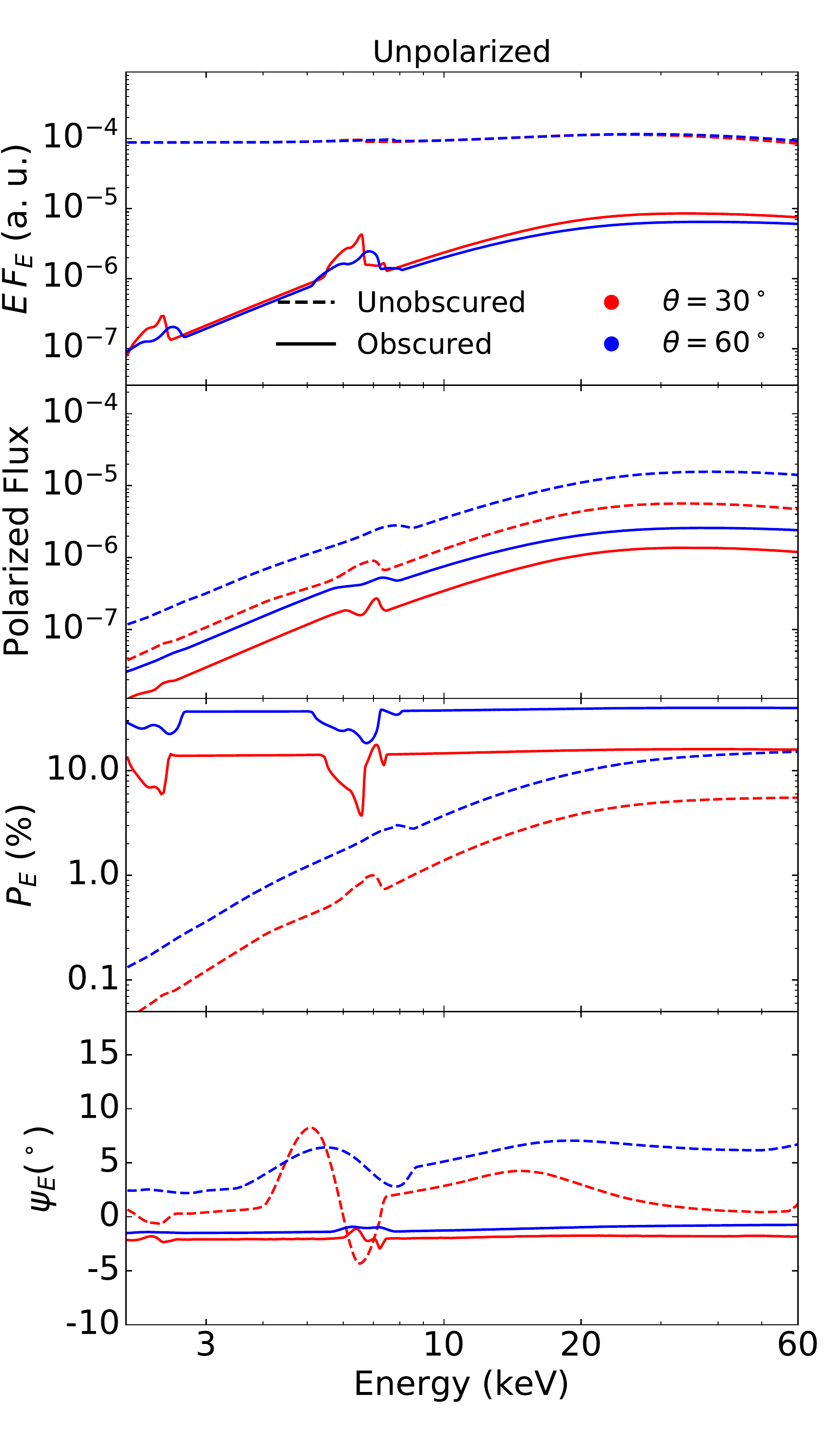}
\includegraphics[width = 0.329\linewidth]{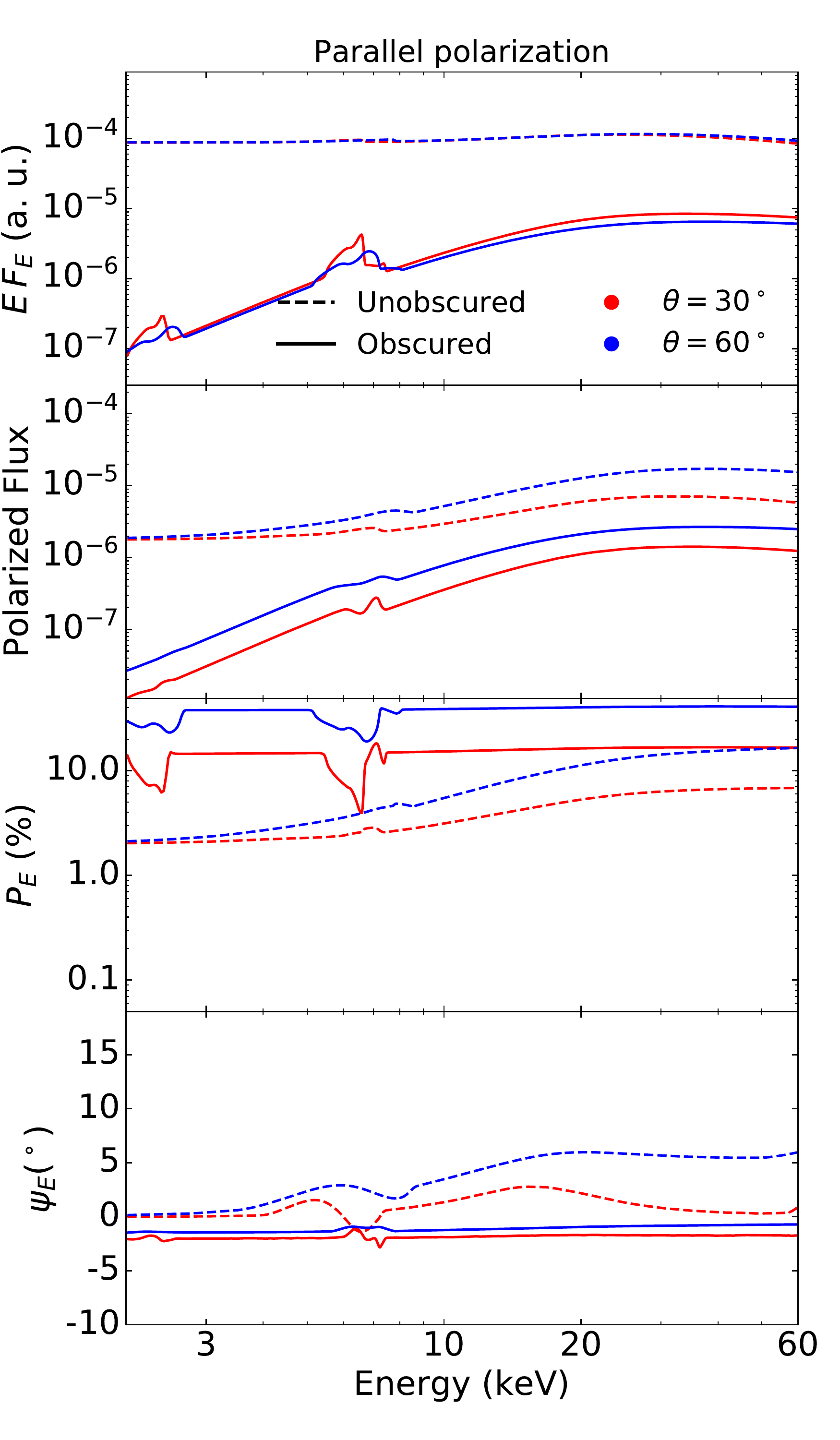}
\includegraphics[width = 0.329\linewidth]{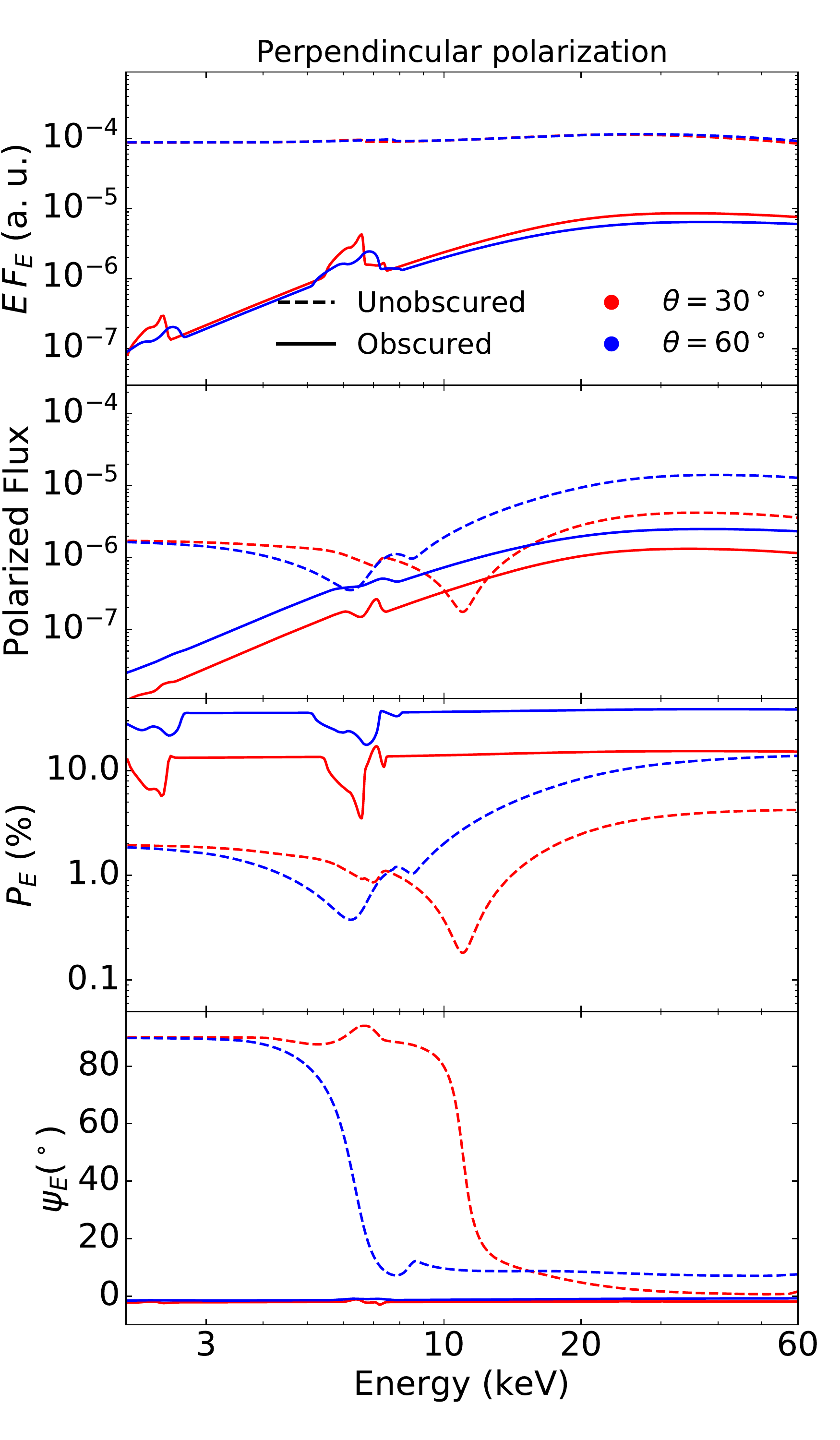}
\includegraphics[width = 0.329\linewidth]{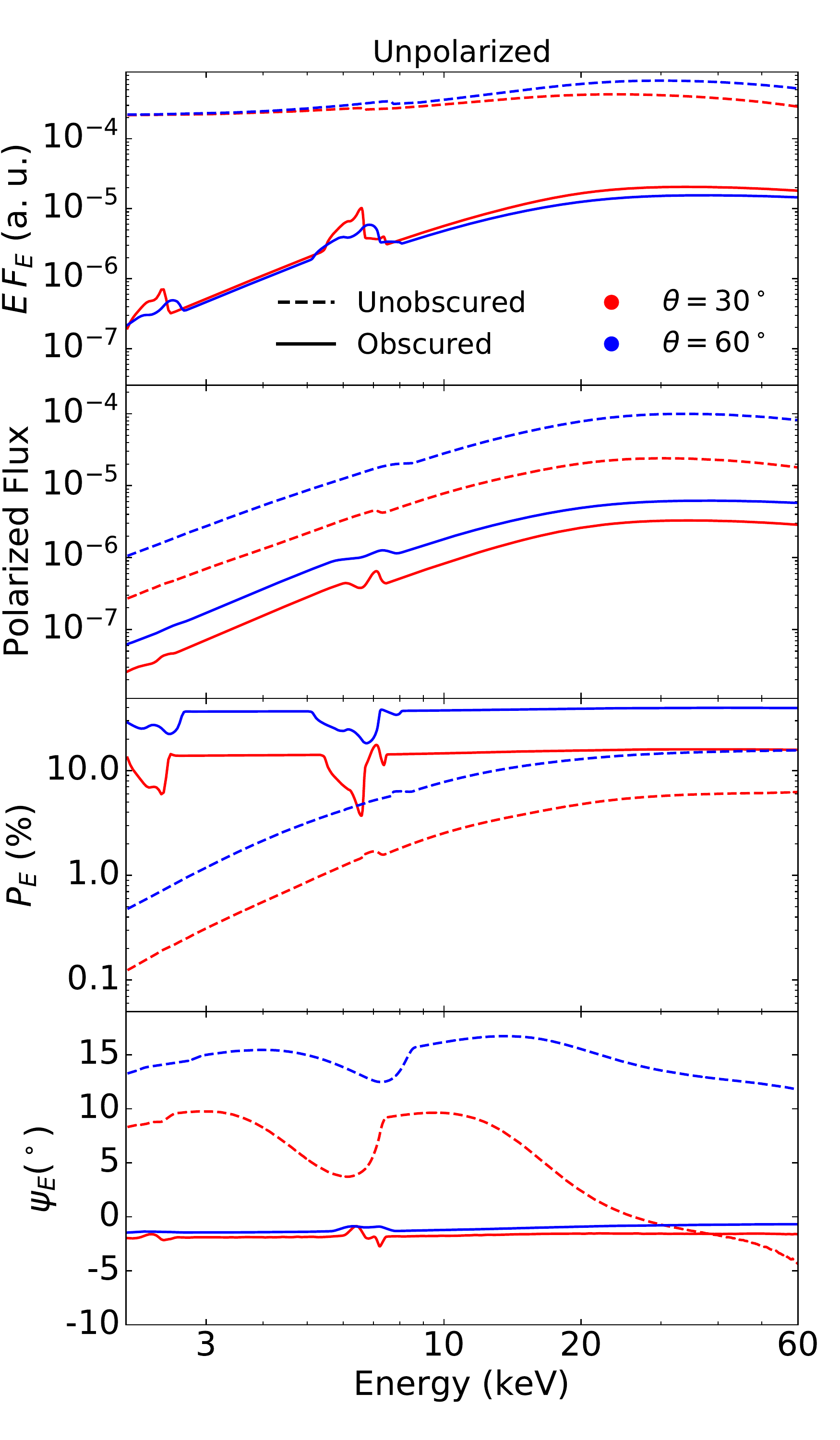}
\includegraphics[width = 0.329\linewidth]{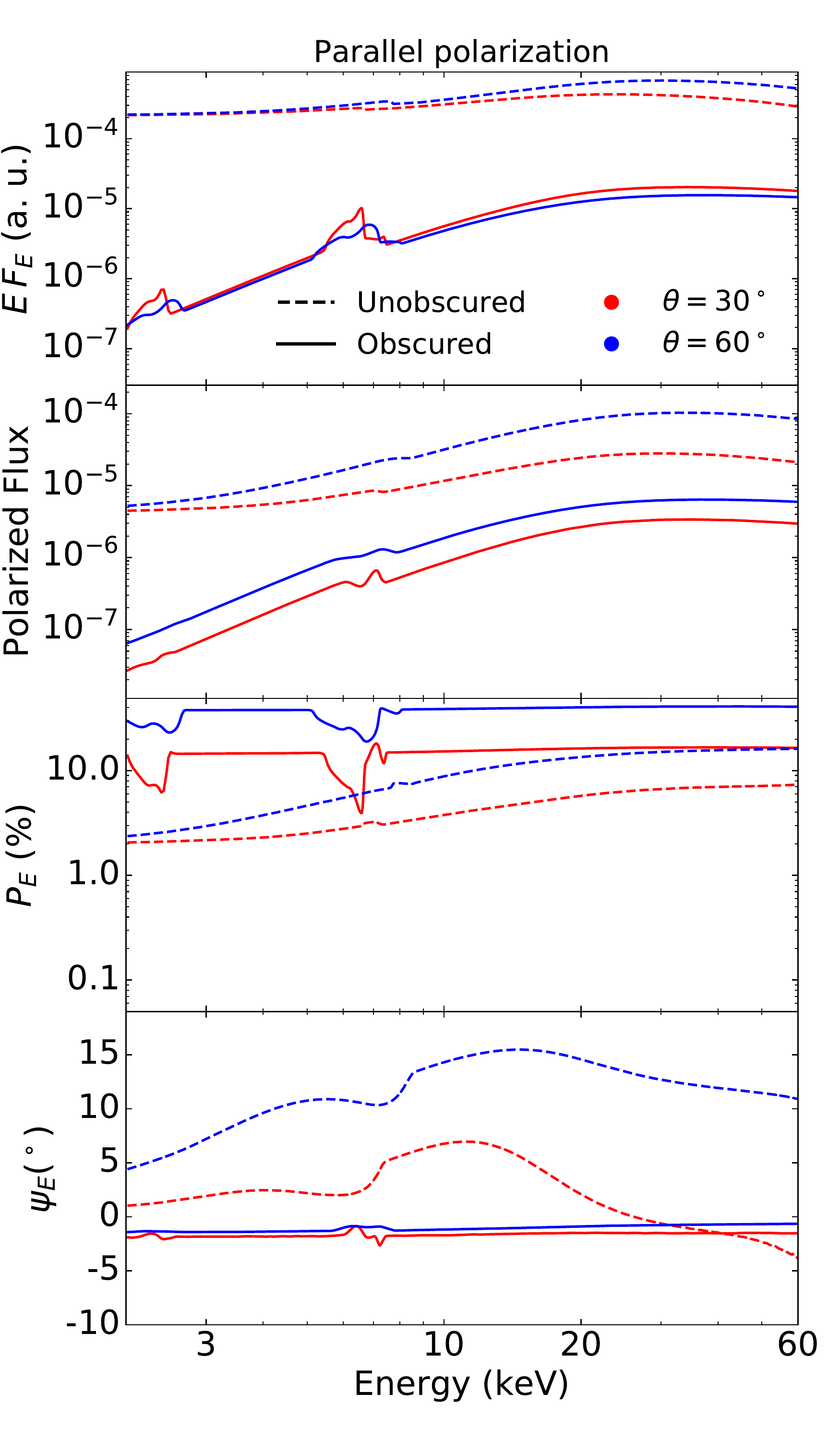}
\includegraphics[width = 0.329\linewidth]{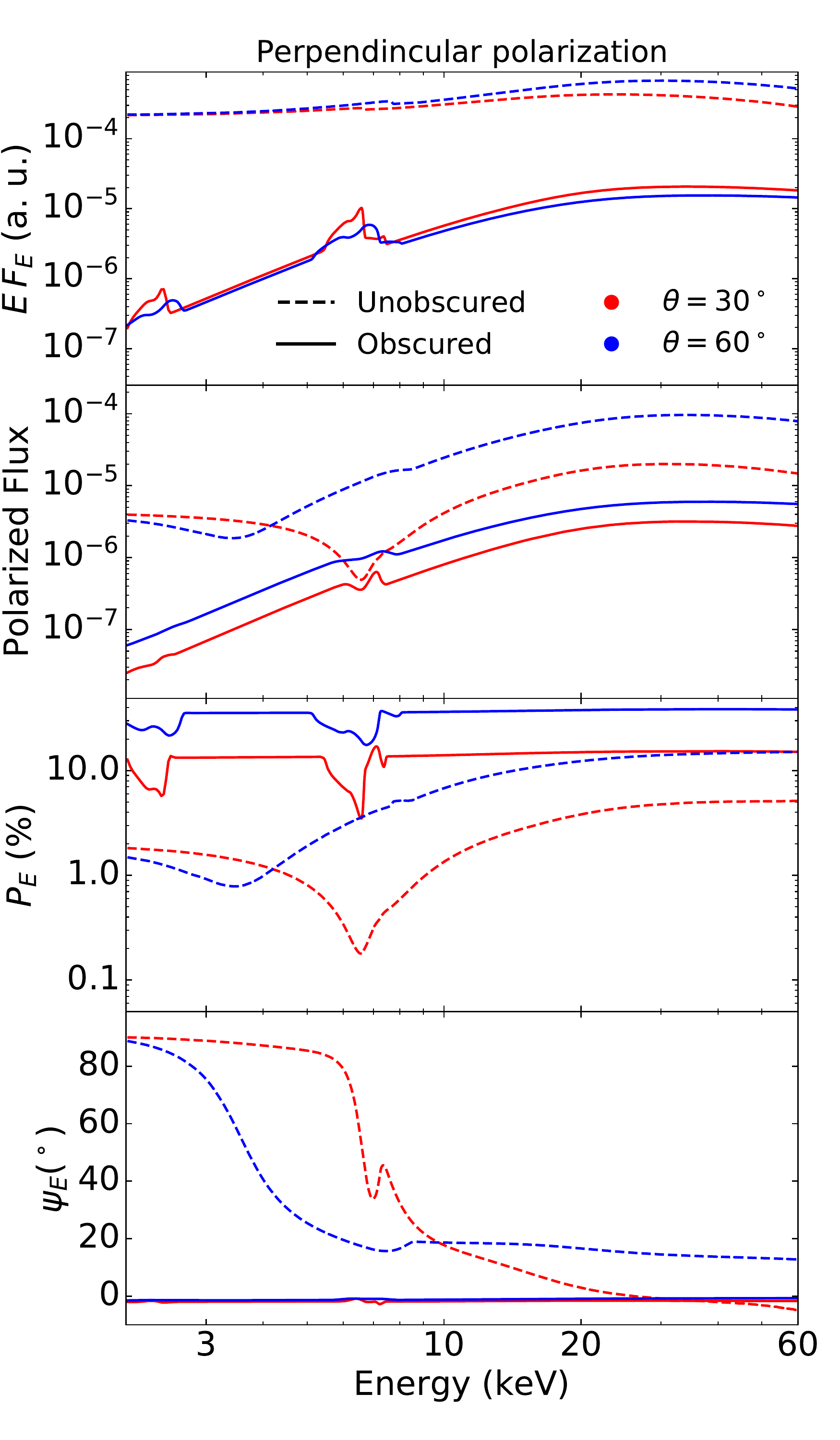}

\caption{X-ray flux ($EF_{E}$), polarized flux ($EF_{E}$ times the 
	 polarization degree), polarization degree $P$ and polarization 
	 position angle $\Psi$ as function of energy seen by an observer 
	 at infinity. We consider a point-source corona at 2.5 $\rm r_g$ 
	 above the BH spinning with $a^\ast = 0$ (top panels) and 
	 $a^\ast = 0.998$ (bottom panels) with an accretion disc that is 
	 inclined by 30\degr\ (red lines) and 60\degr\ (blue lines). The 
	 parameters are shown for two configurations: an unobscured system 
	 (dashed lines) and a system that is obscured by a cloud, of radius
	 30\,$\rm r_g$, that is aligned with the BH.}
\label{fig:Pol_spec}
\end{figure*}


\subsection{Energy dependence}
\label{subsec:Pol_energy}

We present in Fig.\,\ref{fig:Pol_spec} the total flux, the polarized flux defined as the total flux times the polarization degree, the polarization degree and the polarization position angle as a function of energy for unpolarized and polarized primary located at 2.5$\rm r_g$ above the disc. We show the results for both Schwarzschild and Kerr BHs (top and bottom panels, respectively). We considered the cases of an unobscured source and  the case when the primary is obscured by a cloud ($R_{\rm c} = 30\,\rm r_g$) in the line of sight aligned with the BH ($\alpha = 0\,\rm r_g$), as an example. 

\begin{figure*}
\centering
\includegraphics[width = 0.95\linewidth, angle=0]{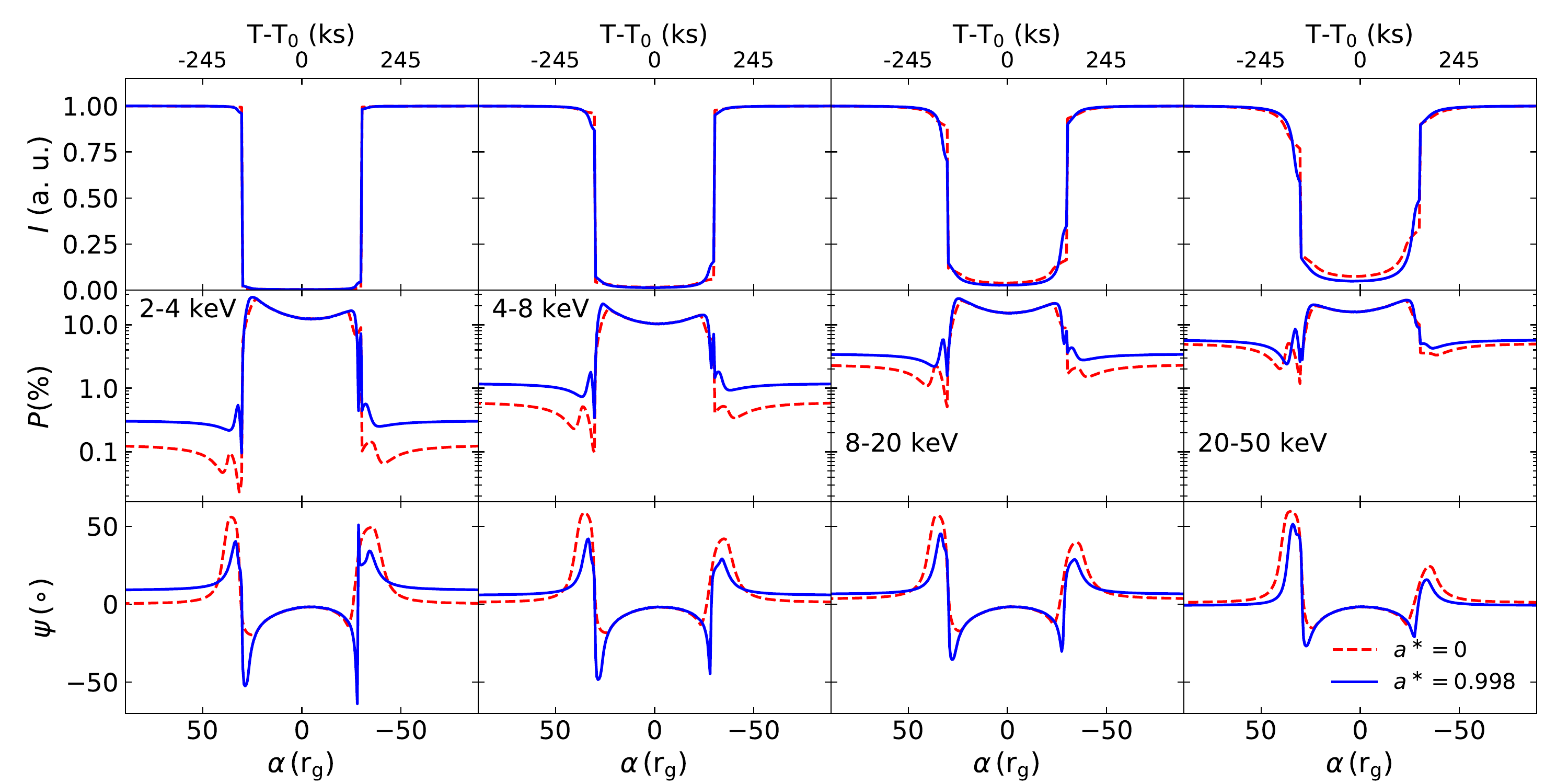}
\includegraphics[width = 0.95\linewidth, angle=0]{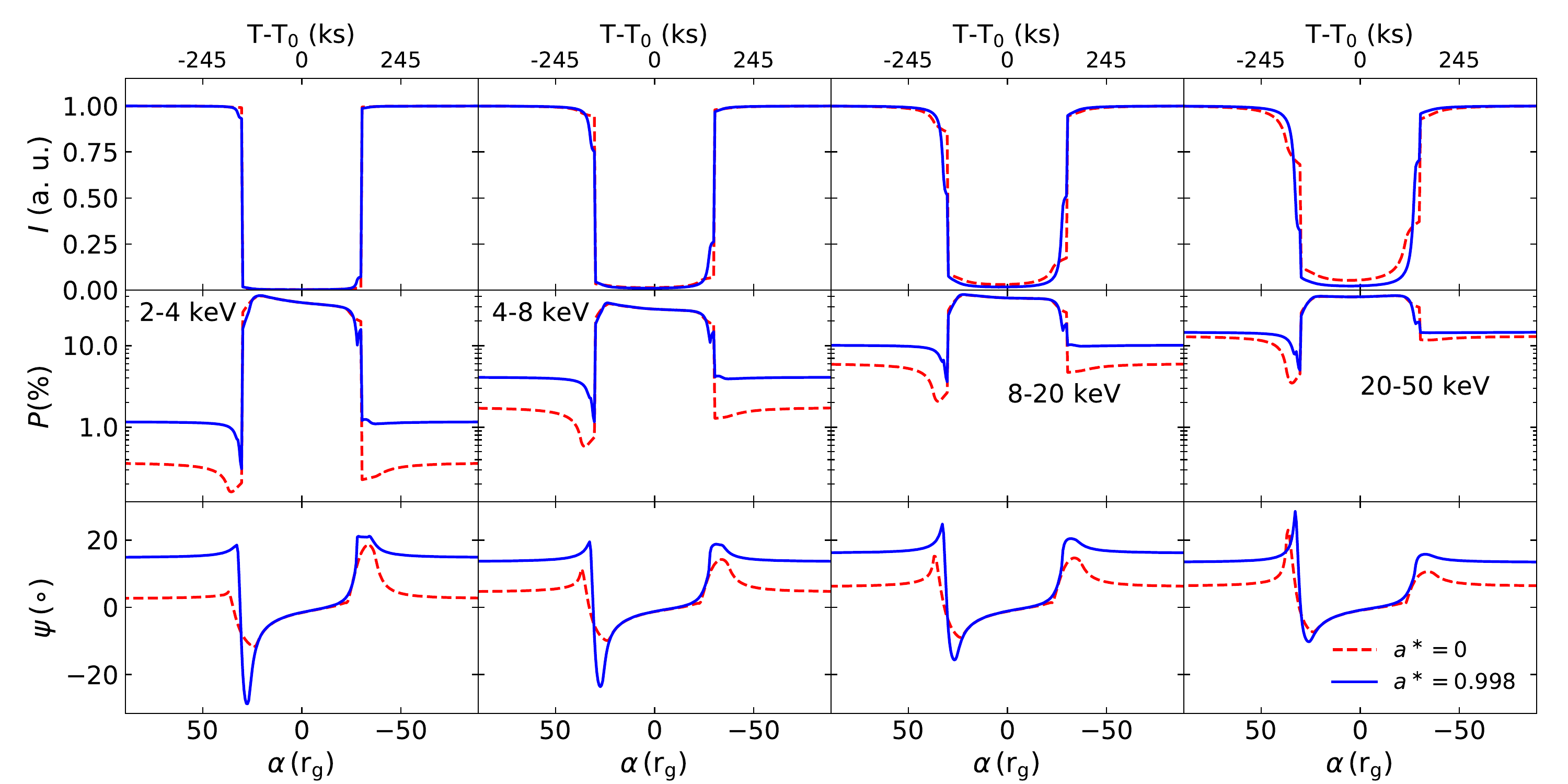}

\caption{Temporal evolution of the X-ray flux, polarization degree ($P$) and polarization position angle ($\Psi$) in the 2-4 keV, 4-8 keV, 8-20 keV and 20-50 keV bands. We considered accretion discs with inclinations of 30\degr\ (top pannel) and  60\degr\ (bottom panel) around Schwarzschild (red lines) and maximally Kerr (blue lines) BHs. We considered an optically thick cloud ($R_{\rm c} = 30\,\rm r_g$) eclipsing the system whose primary ($h=2.5\,\rm r_g$) is unpolarized .}
\label{fig:Pol_LC_Rc30_P0}
\end{figure*}

\begin{figure*}
\centering

\includegraphics[width = 0.95\linewidth, angle=0]{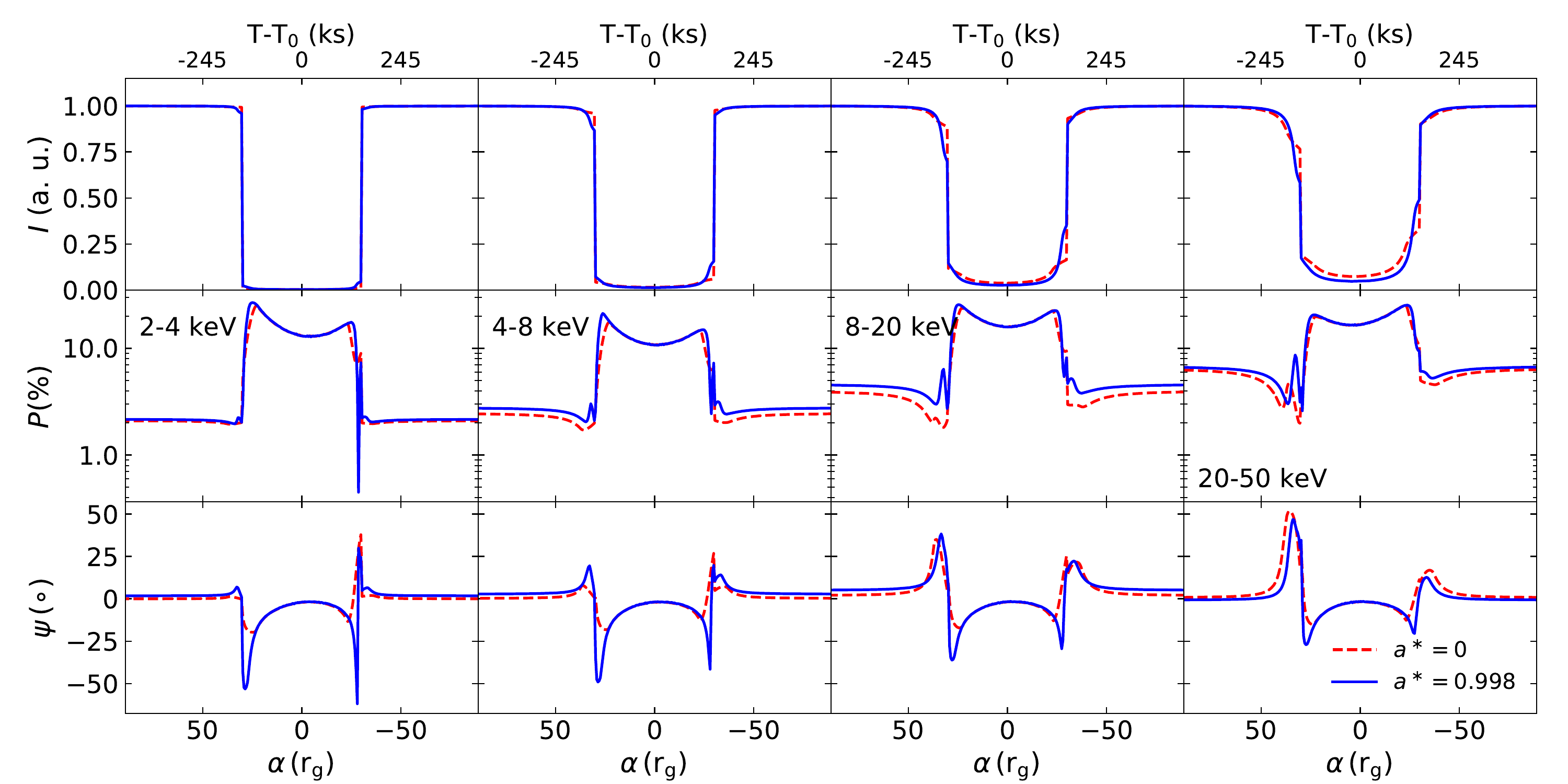}
\includegraphics[width = 0.95\linewidth, angle=0]{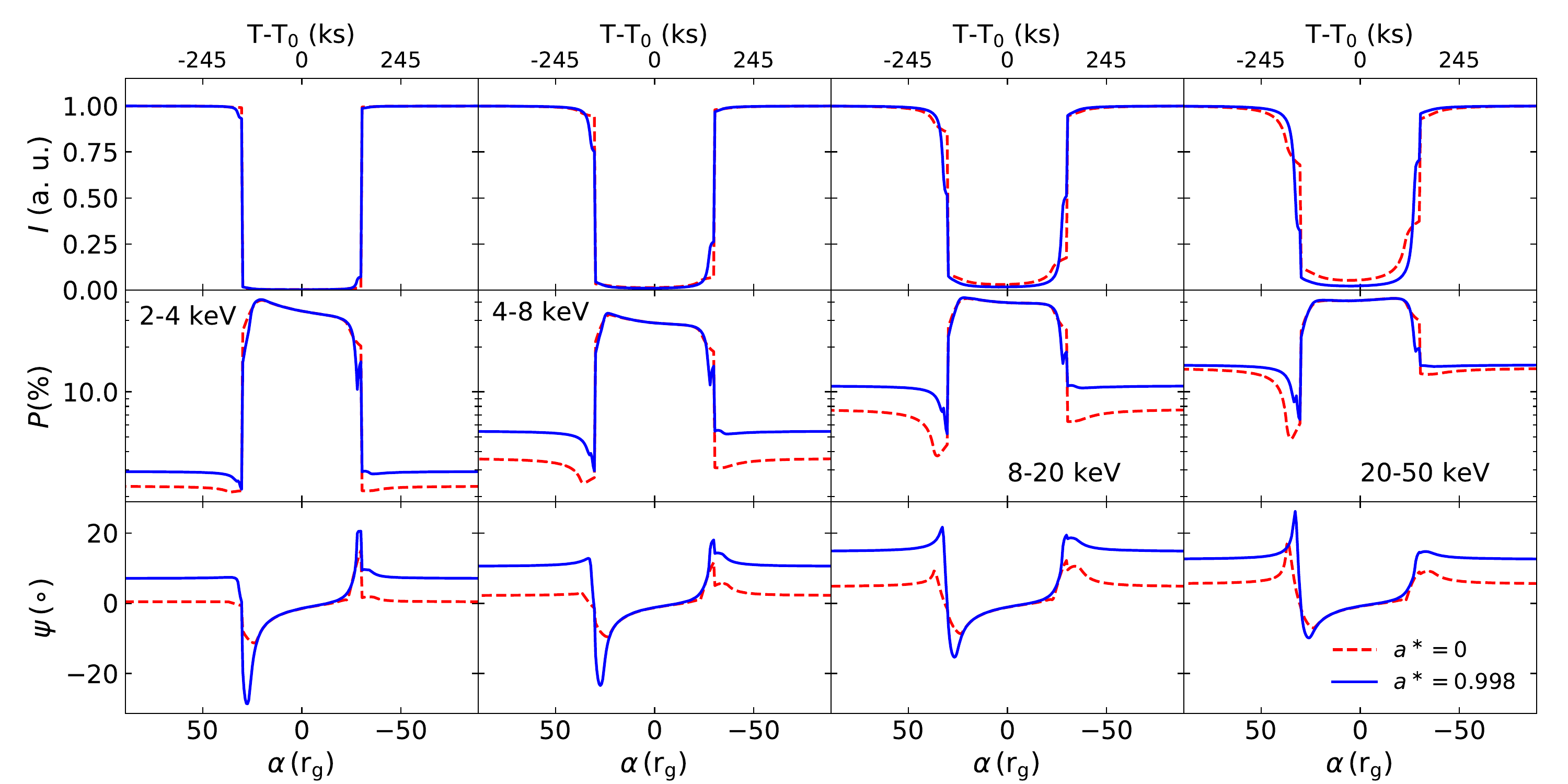}

\caption{Same as Fig.\,\ref{fig:Pol_LC_Rc30_P0} but considering a 2-percent parallely polarized primary.}
\label{fig:Pol_LC_Rc30_P2_Psi0}
\end{figure*}

\begin{figure*}
\centering

\includegraphics[width = 0.95\linewidth, angle=0]{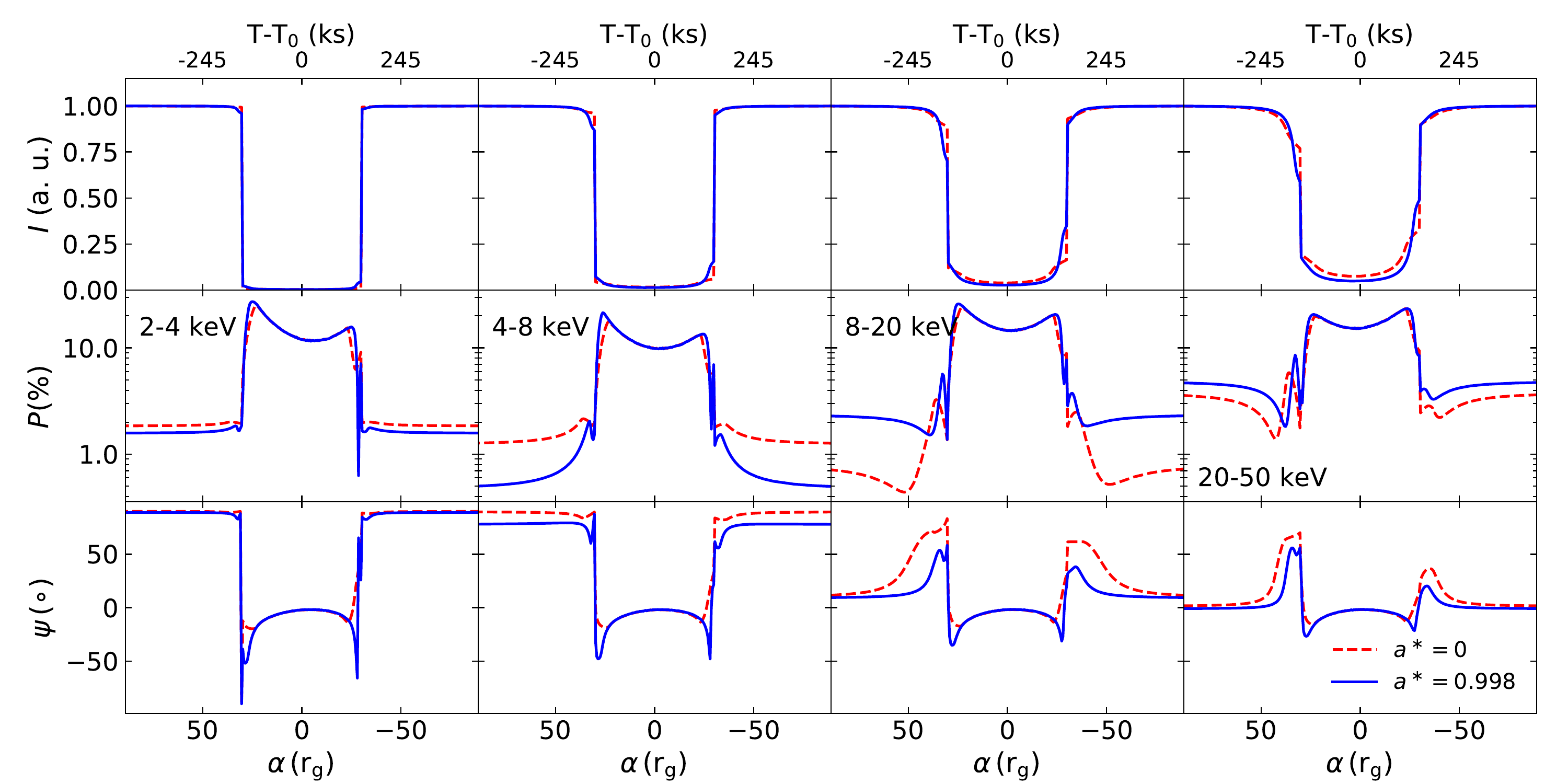}
\includegraphics[width = 0.95\linewidth, angle=0]{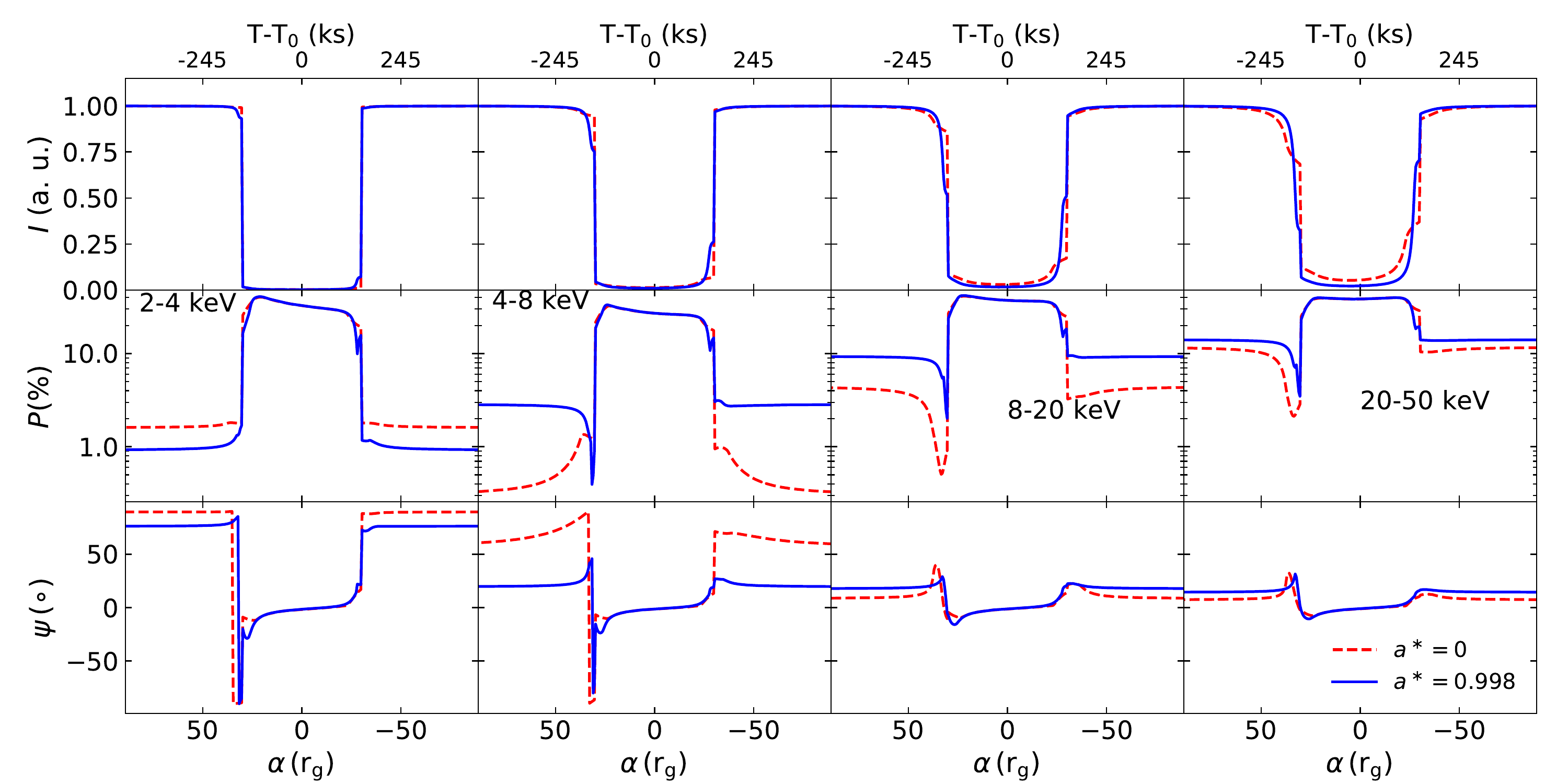}

\caption{Same as Fig.\,\ref{fig:Pol_LC_Rc30_P0} but considering a 2-percent perpendicularly polarized primary.}
\label{fig:Pol_LC_Rc30_P2_Psi90}
\end{figure*}

Considering the unobscured cases, we note that the higher the inclination the higher the polarization degree, due to special and general relativistic effects \citep[][]{Dovciak11}. For the unpolarized and parallelly polarized primary, $P(E)$ increases with energy. At low energies photo-absorption in the disc dominates, thus $P(E)$ is the same as the one of the primary emission. However, for higher energies where the disc reflection becomes more important the polarization varies (increases or decreases) depending on the the polarization angle of the irradiating spectrum. Scattering inside the disc tends to result in parallel polarization angles. If the irradiating emission is parallelly polarized, then the resulting polarization degree increases. Otherwise, the orthogonality between the two vectorial components tends to decrease the net polarization degree. We note that, for the perpendicular polarization, at $\theta =30$\degr,  the polarization degree shows a dip at energies below $\sim 10$\,keV and 7\,keV, for the Schwarzschild and Kerr BHs, respectively. Then it increases continuously at higher energies. A similar behaviour can be seen for $\theta = 60$\degr\, but the transition energies are $\sim 7$ and 4~keV, respectively. These transitions are accompanied by a decrease in the polarization position angle at higher energies, due to the fact that scattering inside the disc becomes more prominent at higher energies, resulting in parallel polarization angles. The energy at which the rotation occurs depends on the spin of the BH, which determines the location of the ISCO. The higher the spin, the closer the ISCO to the BH, which leads to a variation in $\Psi$ at lower energies.

Considering the obscured spectra, which correspond to pure reflection, the polarization degree is boosted up to $\sim 15\%$ and $\sim 38\%$ for $\theta = 30$\degr\ and 60\degr\, respectively. This is due to the obscuration of the primary emission, being either unpolarized or slightly polarized but with a high flux, which tends to dilute the polarization signal from the disc. The $P(E)$ patterns show a clear decrease in polarization due to the broad emission lines from the disc, which are expected to be unpolarized \citep[e.g.][]{Matt93}. Fig.~\ref{fig:Pol_spec} shows a decrease in the polarization position angle for all cases. We stress that the behaviours of $P(E)$ and $\Psi(E)$ are highly dependent on the position of the cloud as it eclipses different parts of the accretion disc, as we will discuss in the next section.

\subsection{Time dependence}
\label{subsec:Pol_time}


{We explore in this section the variability (on a timescale up to a few hundred ks)} that is introduced in the polarization signal due to obscuration events. Figs.~\ref{fig:Pol_LC_Rc30_P0}-\ref{fig:Pol_LC_Rc30_P2_Psi90} show the temporal evolution of the flux, the polarization degree and polarization position angle as a cloud of radius $30\,\rm r_g$ eclipses the innermost parts of an AGN. We present in Fig.~\ref{fig:Pol_LC_Rc30_P0} the results assuming an unpolarized primary located at 2.5\,$\rm r_g$ above either a Schwarzschild or a Kerr BH, for inclinations of 30\degr\ and 60\degr. In Figs.~\ref{fig:Pol_LC_Rc30_P2_Psi0}~-\ref{fig:Pol_LC_Rc30_P2_Psi90}, we present the same results for a 2 percent (parallelly and perpendicularly, respectively) polarized primary. We considered 4 energy bands: 2--4, 4--8, 8--20 and 20--50\,keV. The polarization signal of the accretion disc is not uniform due to relativistic effects \citep[see e.g.][]{Schnittman09, Schnittman10}, as shown in Fig.~\ref{fig:polflux}. Thus the variability pattern seen in these figures is caused by receiving signal from different patches of the disc as the cloud is moving across the line of sight.

We consider first the unpolarized primary case for an inclination of 30\degr. In the 2--4~keV band, the polarization degree starts from a negligible value in the unobscured case then it shows a small increase, followed by a decrease, as the cloud is moving closer to the center ($\alpha \simeq 36$ and 32~$\rm r_g$ for the Schwarzschild and Kerr BHs, respectively) that is accompanied by a little decrease in flux and a remarkable variation in polarization position angle. This effect is mainly due to the obscuration of the depolarizing region located to the North-West of the BH (see Fig.\,\ref{fig:polflux}). We note that this effect is smaller for higher inclinations and/or low spin values since, in these cases, this region is very close to the horizon. As the cloud moves further it eclipses the lamp-post causing the flux to drop drastically and increasing $P$, which reaches $\sim$25\% (at $\alpha \simeq 23\,\rm r_g$) and 26\% (at $\alpha \simeq 25\,\rm r_g$) for Schwarzschild and Kerr BHs, respectively. However, $\Psi$ rotates to negative values of $\sim -25$\degr\ and $\sim -50$\degr\ for Schwarzschild and Kerr BHs, respectively. $P(t)$ decreases with time and reaches $\sim 12$\% at $\alpha = 0\,\rm r_g$, for both spins, while $\Psi(t)$ increases up to $\sim -1$\degr. As the cloud is moving away from the center, $P(t)$ increases again up to $\sim 16$\% at $\alpha \simeq -23\,\rm r_g$ and $-25\,\rm r_g$ for Schwarzschild and Kerr BHs, respectively. This is accompanied by a decrease in $\Psi(t)$. As the cloud is moving further out and uncovering the lamp-post $P(t)$ decreases and re-increases suddenly at $\alpha \simeq -30\,\rm r_g$ ($\Delta P \simeq 2\%$ and 6.7\% for Schwarzschild and Kerr BHs, respectively). As for $\Psi(t)$, its behaviour is inverted for the Kerr BH, while it increases continuously for the Schwarzschild case. This effect is due to unobscuring the depolarizing region described above. We also note that the approaching part of the disc contributes substantially to the polarization degree due to Doppler boosting (as shown in Fig.\,\ref{fig:polflux}). This effect becomes more important for higher inclinations. Therefore, obscuring/unobscuring this part of the disc will lead to a decrease/increase in $P(t)$ before obscuring/unobscuring the primary source. As the cloud is moving out of the line of sight all the quantities go back smoothly to their original values. Qualitatively, a similar variability pattern can be seen for all energy bands, having different values as various processes and spectral features dominate in different energy bands. For example, the increase in polarization degree is the smallest in the 4--8~keV band, which is affected by the presence of the broad unpolarized Fe line. In general, the variation of $P(t)$ can be described by an asymmetric double peaked profile, where the peaks correspond to the start and the end of the eclipse, being larger for the former case. However, we note that this is inverted for the 20--50~keV band where the second peak is higher. We note that $P(t)$ and $\Psi(t)$ show a qualitatively similar behaviour for the parallelly polarized primary as well.

\begin{figure*}
\centering
\includegraphics[width = 0.32\linewidth]{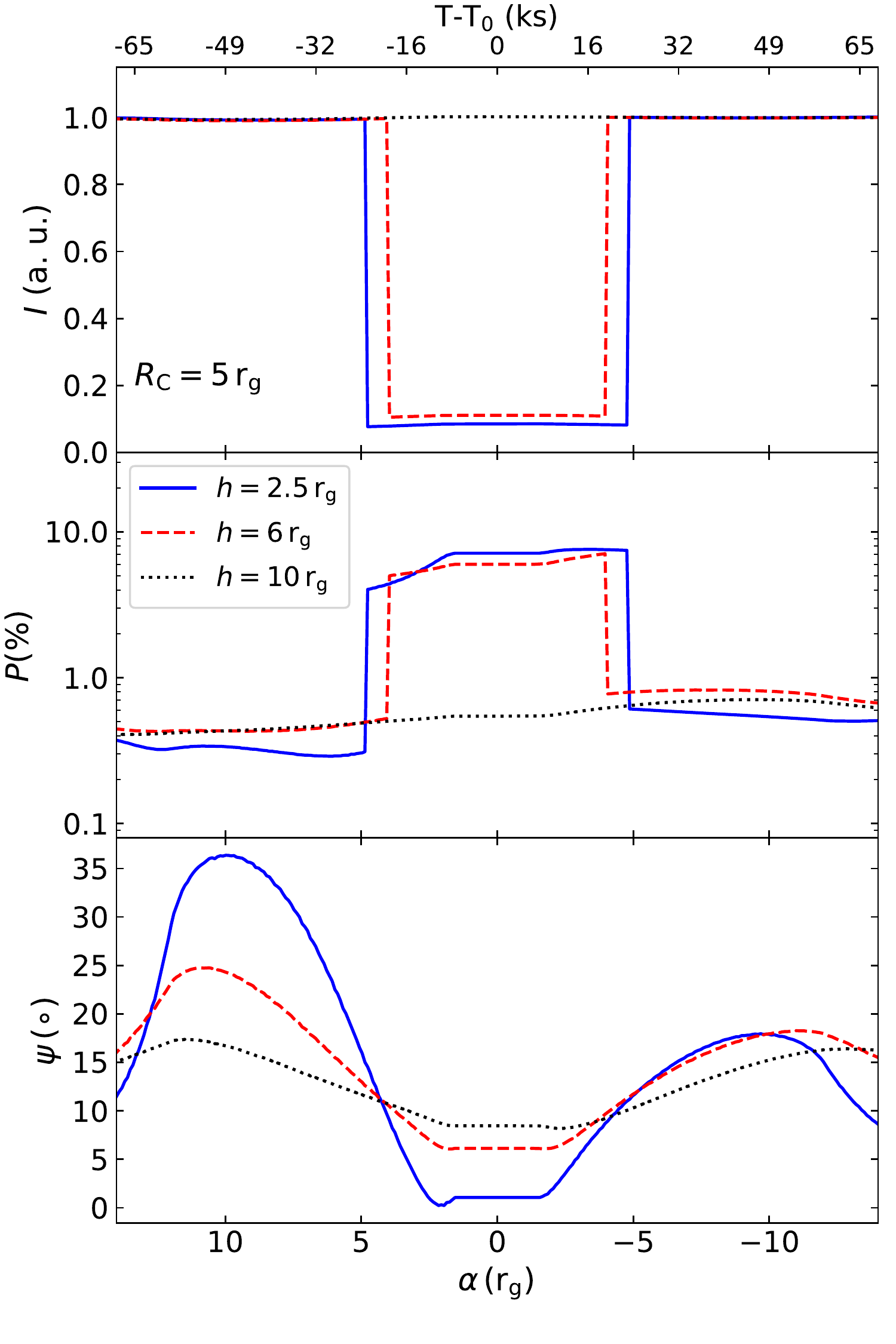}
\includegraphics[width = 0.32\linewidth]{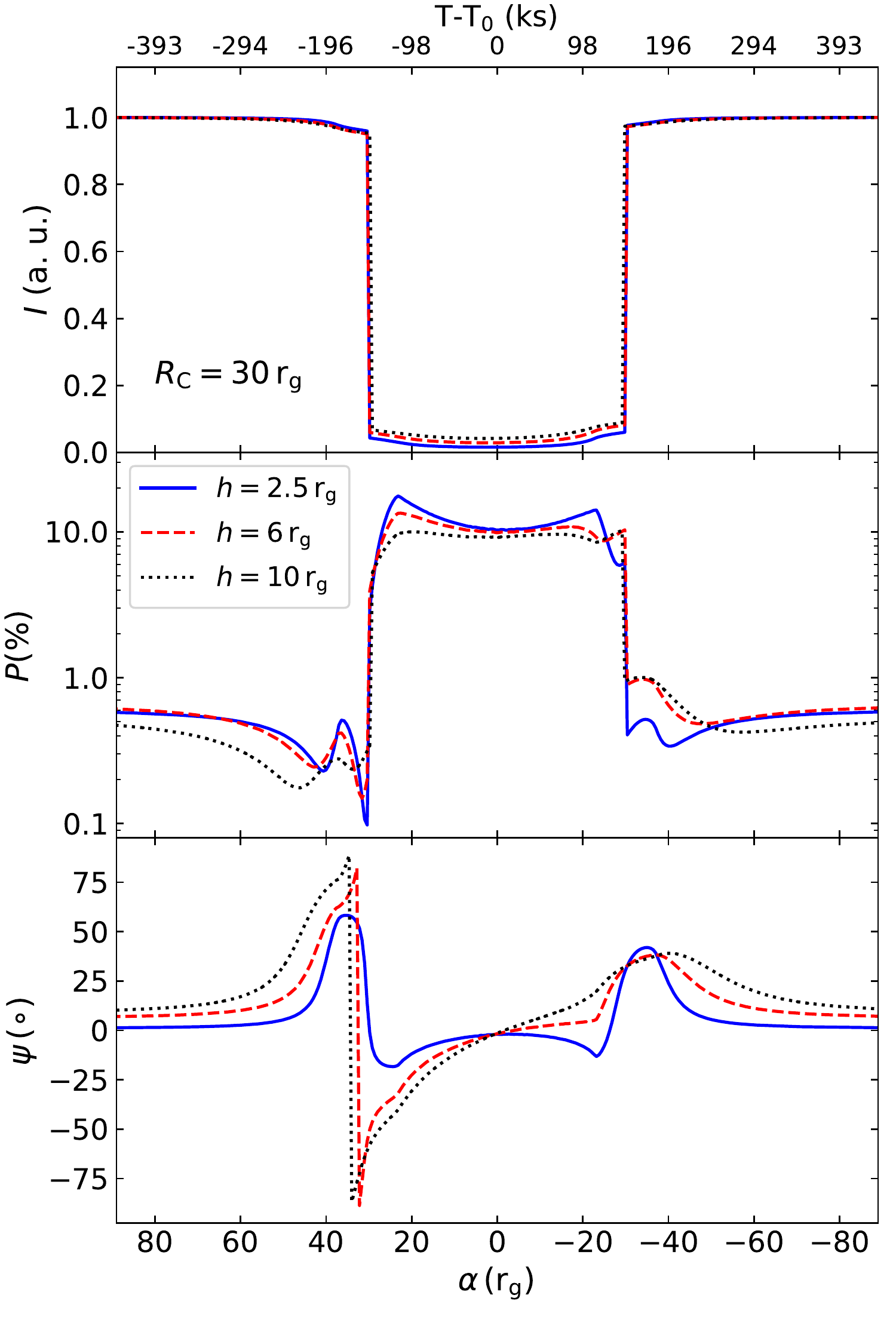}
\includegraphics[width = 0.32\linewidth]{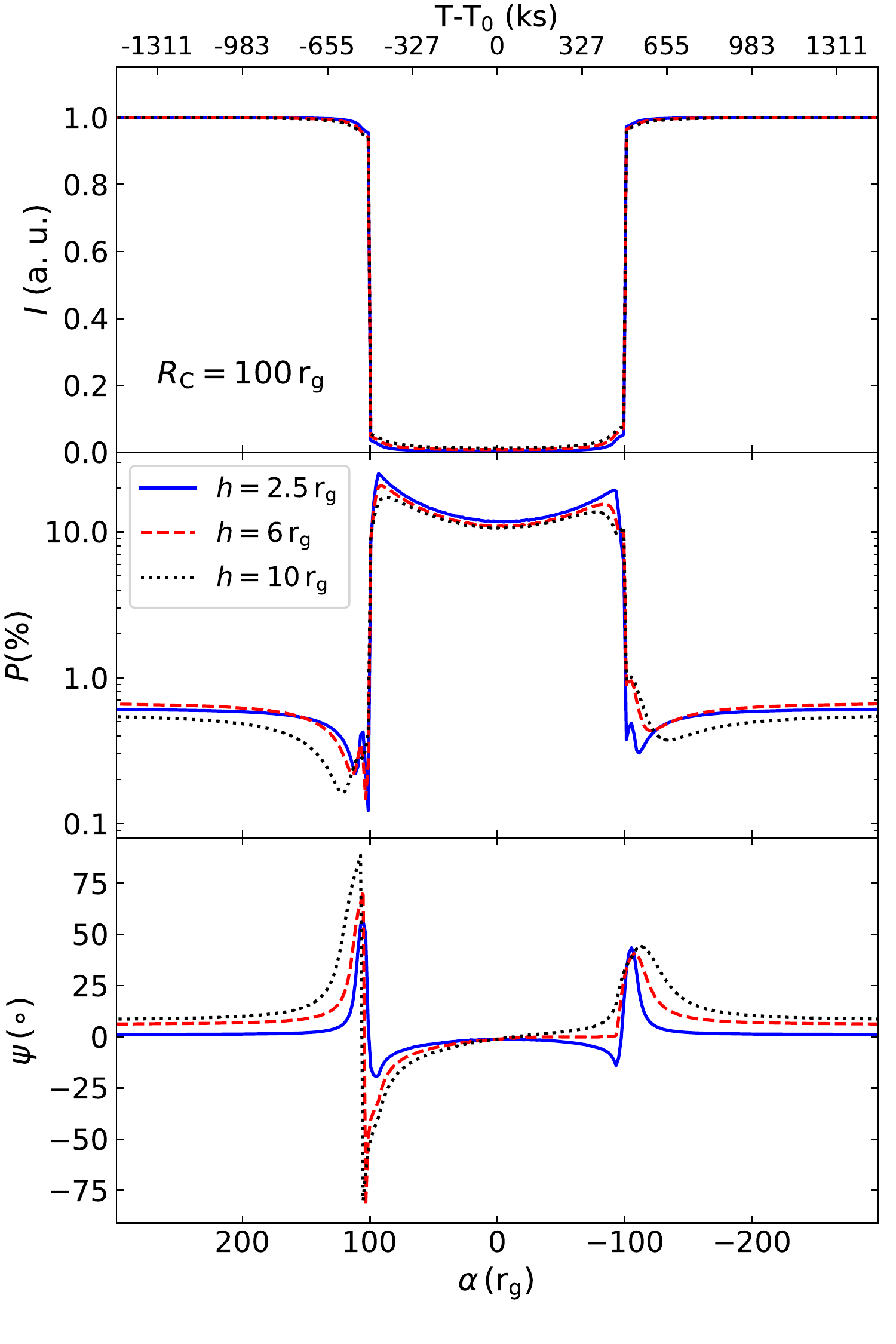}\\
\includegraphics[width = 0.32\linewidth]{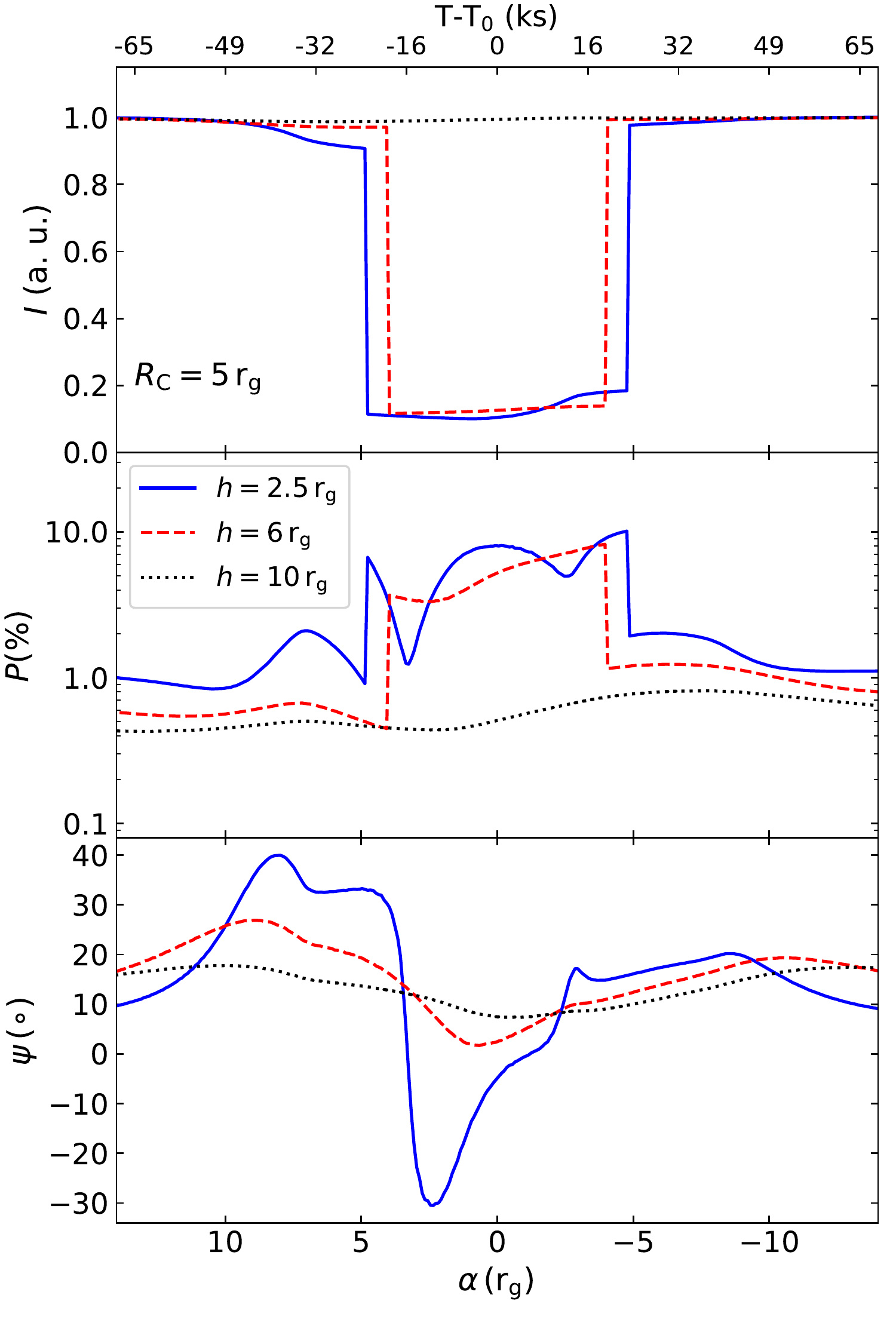}
\includegraphics[width = 0.32\linewidth]{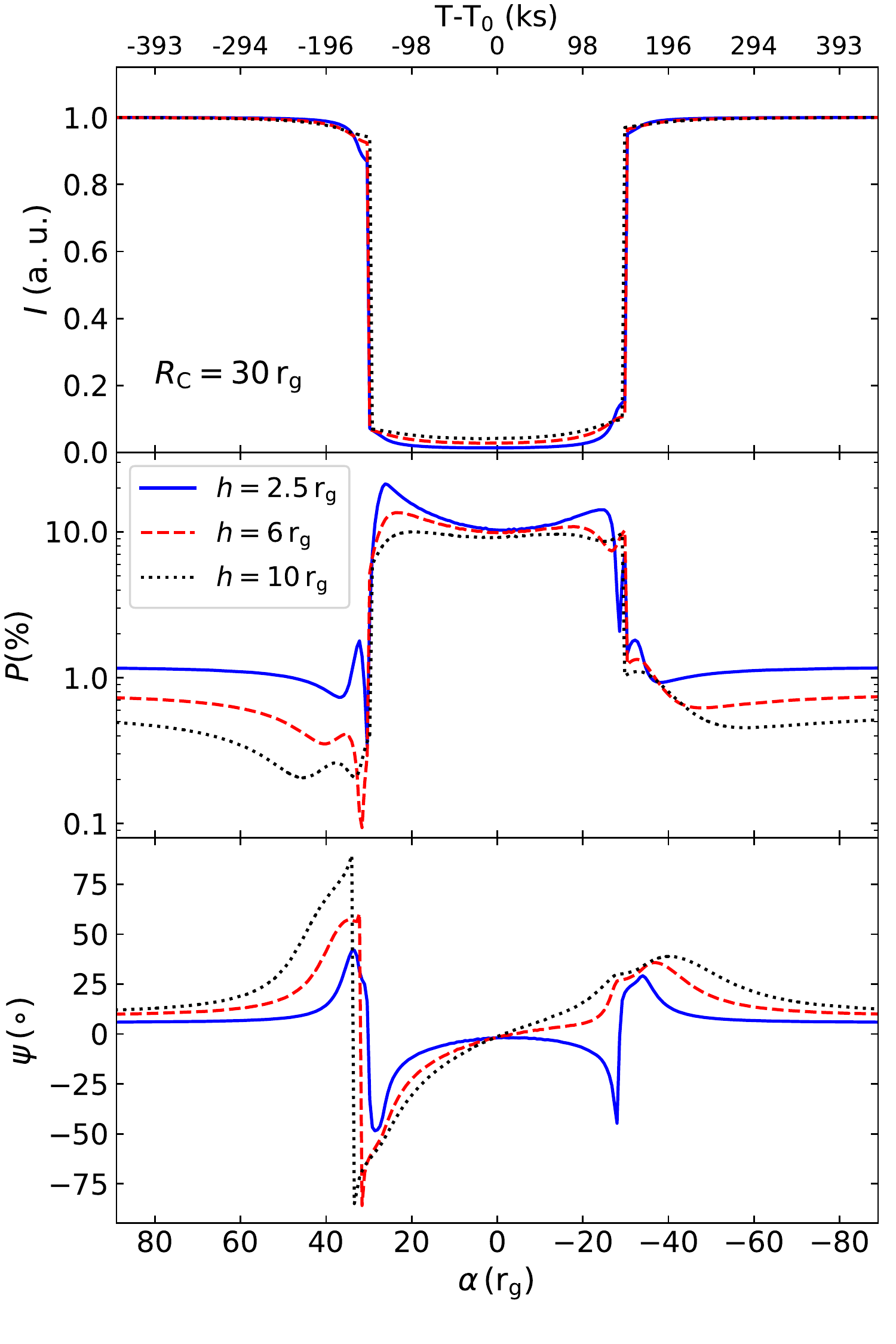}
\includegraphics[width = 0.32\linewidth]{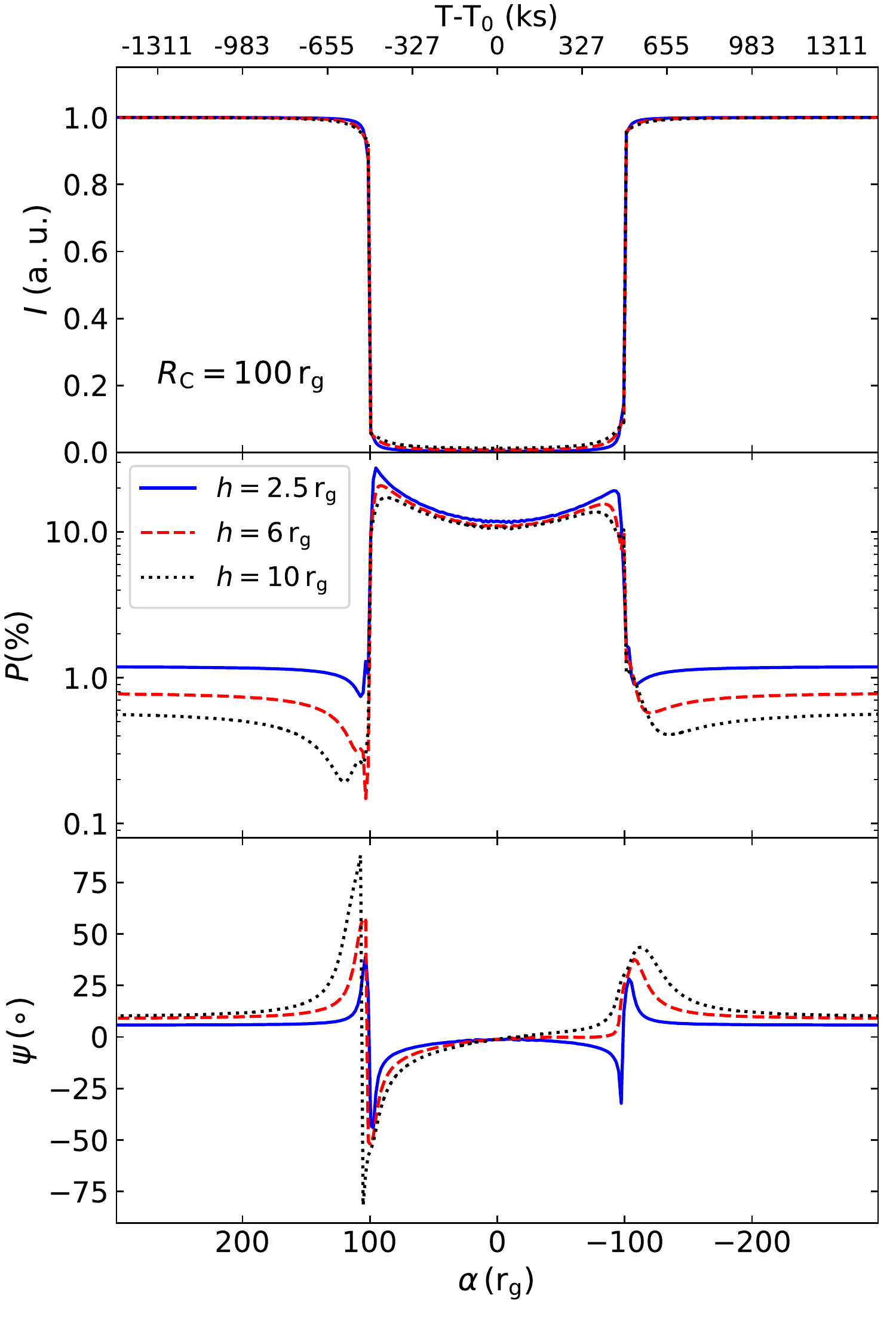}
\caption{Temporal variability of the flux, polarization degree ($P$) and polarization position angle ($\Psi$) in the 4--8 keV band for an unpolarized primary at $h=2.5\,\rm r_g$ (solid blue lines), $h=6\,\rm r_g$ (dashed red lines) and $h=10\,\rm r_g$ (dotted black lines), assuming a BH with $a^\ast = 0, 0.998$ (top and bottom panels, respectively) and an inclination $\theta = 30$\degr. We considered a cloud of radii 5, 30 and 100 $\rm r_g$ (left, middle, right columns, respectively).}
\label{fig:Pol_compareH}
\end{figure*}


For a perpendicularly polarized primary, $P(t)$ shows a similar behaviour compared to the other two cases. However, $\Psi(t)$ in the 2--4~keV and 4--8~keV bands starts from $\sim 90$\degr\ when the lamp-post is unobscured then rotates quickly to negative values as soon as the primary is obscured, showing then a variability pattern similar to the parallelly polarized and unpolarized primary. $\Psi(t)$ then increases again to $\sim 90$\degr\ when the cloud is moving away from the line of sight. However, at higher energies, where scattering in the disc is more prominent, leading to a more parallel polarization, $\Psi(t)$ varies in a similar fashion to the previous polarization scenarios. We note that in all cases, both Schwarzschild and Kerr BHs show the same pattern when the innermost regions of the disc are obscured, while the patterns differ once these regions are unobscured. This is due to the fact that in the case of a Kerr BH the accretion disc reaches lower radii with respect to a Schwarzschild BH giving different observed polarization states.

We consider now the case of a higher inclination of $\theta =60$\degr. $P(t)$ varies in a similar way to the $\theta = 30$\degr\ case described above. However, we note that on the one hand it reaches higher values during obscuration, $\sim 40\%$. On the other hand, the double peaked features are smoother. As for $\Psi(t)$, we note that the differences between the Schwarzschild and Kerr BHs, for the uncovered primary, are bigger compared to the case of $\theta = 30$\degr, being $\sim 2$\degr\ and $15$\degr, respectively, in the 2--4~keV band (unpolarized and parallelly polarized scenarios). Then $\Psi$ drops to lower values as the primary source is covered and rises again while the cloud is moving towards the receding part of the disc, to reach its initial value when the cloud moves away from the line of sight. Similarly to the lower inclination case, when the source is not obscured, $\Psi(t)$ is larger for energies below 8~keV compared to the one at higher energies.

\subsection{Effects of the lamp-post height and the cloud radius}
\label{subsec:Pol_h_Rc}

We investigate in Fig.\,\ref{fig:Pol_compareH} the effects of the height of the primary as well as the size of the cloud on the polarization signal. This figure shows the temporal variability in the 4--8\,keV range for a non-rotating and a maximally rotating Kerr BH (top and bottom panels, respectively), for an unpolarized primary source located at 2.5, 6 and 10\,$\rm r_g$ above the disc ($\theta = 30\degr$). We considered cloud radii of 5, 30 and 100~$\rm r_g$ (for $N_{\rm H} = 10^{24}\,\rm cm^{-2}$, the number density of the clouds will be $\sim 10^{11},\, 10^{10}$ and $10^9\, \rm cm^{-3}$, respectively, covering the whole range of BLR cloud densities). The first immediate difference between the various cloud sizes is the duration of the eclipse which is proportional to the cloud radius.  

For $R_{\rm C} = 5 \, \rm r_g$, the obscured regions of the disc are smaller than the other cases. We note that the eclipse occurs earlier when the source is at 2.5~$\rm r_g$ compared to the 6~$\rm r_g$ case. For these two cases, the variations of $P(t)$ while the cloud is moving within the line of sight are smoother for $a^\ast = 0$. This is due to the fact that the disc does not extend down to the regions, shaded by the cloud, where the effects should be the most intense. However, a larger gradual and asymmetric variability of $\Psi$ is expected, during the motion of the cloud, being $\sim 35\degr$\ for $\alpha = 10~\rm r_g$ and decreasing to 1\degr when the cloud is eclipsing the source, then rising up to $\sim 18\degr$\ for $\alpha = -10~\rm r_g$, for $h=2.5~\rm r_g$. More features are expected for the rotating BH case, especially for $h=2.5~\rm r_g$ as the ISCO shrinks to lower radii. In these cases, $P(t)$ and $\Psi(t)$ show more variability for low heights. We note that for $a^\ast = 0.998$, $P(t) \simeq 1\%$ for the unobscured-primary case while it is $\sim 0.3\%$ for $a^\ast = 0$. For $h=10~\rm r_g$, despite the fact that the source is not eclipsed at all some small variations are still expected in the polarization signal due to the obscuration of various parts of the accretion disc.

As for $R_{\rm C} = 30$ and $100~\rm r_g$, $P(t)$ shows qualitatively a similar variability pattern for the various heights, which is also consistent with the one described in \S\,\ref{subsec:Pol_time}. However, the patterns get smoother and flatter for larger heights, during the eclipse event. We also note that the polarization degree is higher for lower lamp-post heights, as already discussed by \cite{Dovciak11}. In fact, during the unobscured phases $P(t)\simeq 1.2\%,\, 0.7\%$ and 0.5\% for $h = 2.5$, 6 and 10\,$\rm r_g$, respectively. Once the primary is obscured, these values reach a maximum of $\sim 21\%$, 13.5\% and 10\%, respectively, for $R_{\rm C} = 30 ~\rm r_g$ and $a^\ast = 0.998$. The difference in $P(t)$ for different heights is smaller for the Schwarzschild BH case, $P(t) \simeq 0.6\%$ for $h = 2.5$ and $6~\rm r_g$ and 0.4\% for $h = 10~\rm r_g$. $\Psi(t)$ shows a similar variability pattern for the three heights, however the transitions (rise/decay) are more gradual at larger heights (when $\alpha \gtrsim 30,\, 100~\rm r_g$, for $R_{\rm C} = 30$ and $100~\rm r_g$, respectively). Instead, for larger heights, $\Psi(t)$ reaches higher values at the beginning of the eclipse, being $\sim 42$\degr, 60\degr, and 90\degr\ for $h=2.5$, 6 and 10\,$\rm r_g$, respectively at $\alpha \simeq 35\,\rm r_g$, for $R_{\rm C} = 30~\rm r_g$.

\subsection{Effects of parsec-scale material}
\label{subsec:Pol_wind_torus}

In practice the central SMBH, its accretion disc and corona are not isolated from the other AGN components. The impact of absorption, re-emission and scattering of parsec-scale structures has been explored in detail for spectroscopic and photometric purposes \citep[see e.g.][]{Ghisellini94a}. Higher torus hydrogen column density will impact on the Compton hump \citep{Murphy09, Murphy11}. The impact of additional reprocessing from parsec-scale AGN structures on the X-ray polarization signal from type-1 AGN has been neglected until recently. In \cite{Marin18b}, the authors investigate the differences between an isolated central engine and a real type-1 AGN, accounting for biconical narrow line region (NLR) and an equatorial obscuring region. Among the main conclusions, it was found that additional parsec-scale scattering increases the expected polarization degree by $0.5 - 1$ percentage points in the case of an unpolarized or a parallelly polarized continuum source. Changes are more profound for a perpendicularly polarized primary.

\begin{figure}
\centering
\includegraphics[width = 0.49\textwidth]{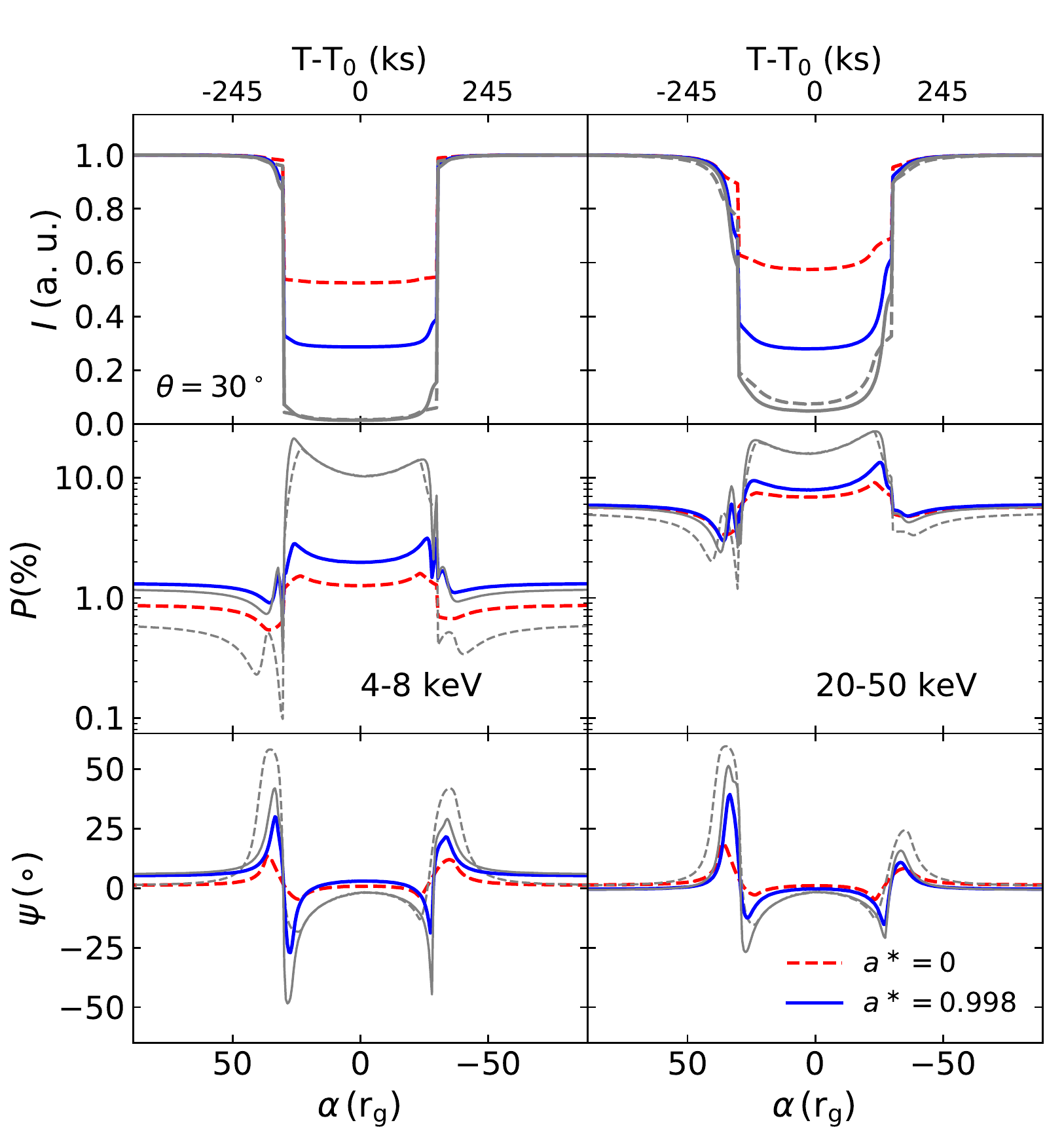}\\
\includegraphics[width = 0.49\textwidth]{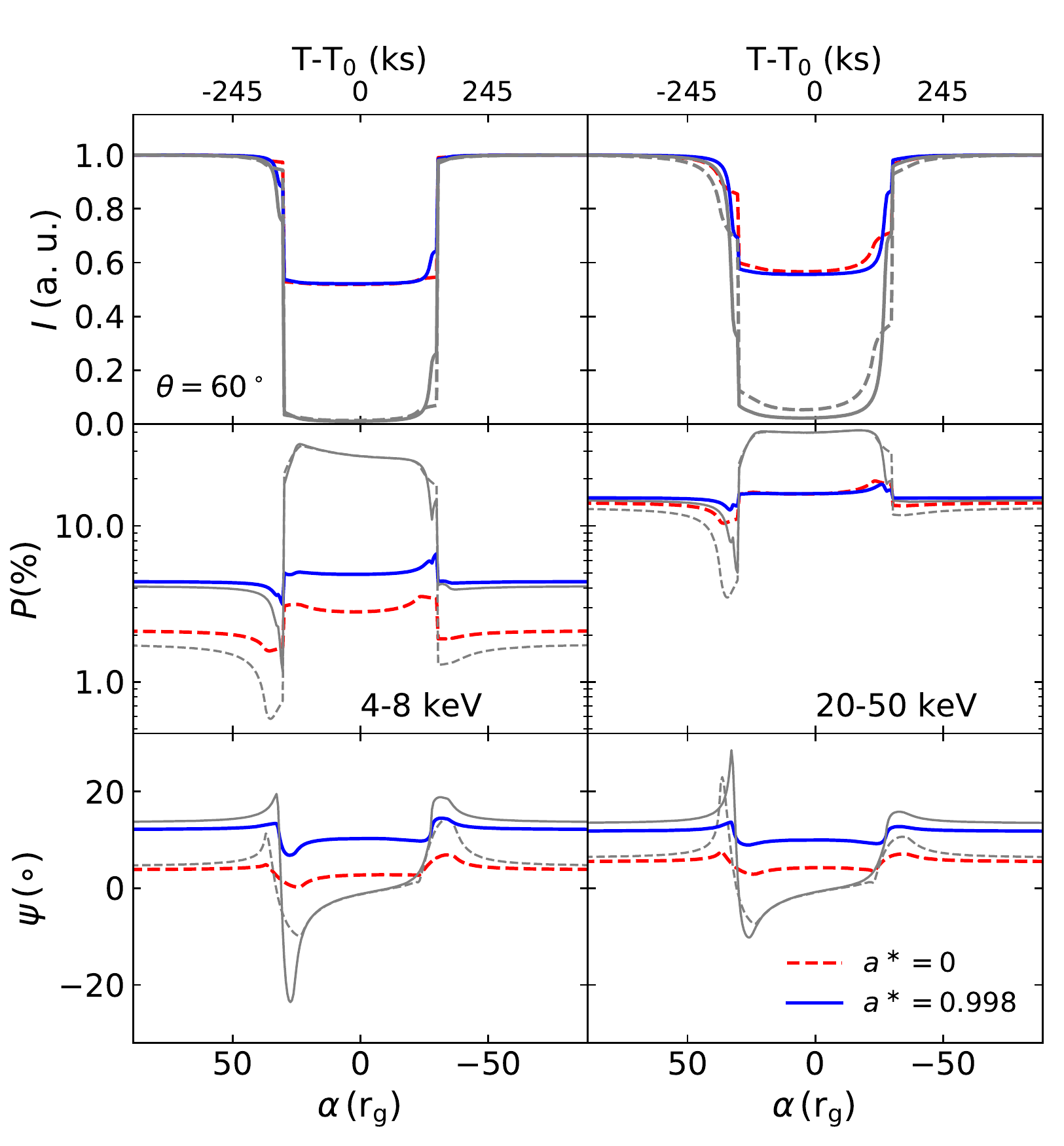}
\caption{Similar to Fig.\,\ref{fig:Pol_LC_Rc30_P0} but for 4--8 keV and 20-50 keV ranges only, taking into account the contribution from the bi-conical narrow line regions and molecular torus for inclinations of 30\degr\ (top panel) and 60\degr\ (bottom panel), for Schwarzschild (dashed lines) and Kerr (solid lines) BHs. The grey lines represent the temporal variations of the polarization degree and polarization position angle without including the effects of the NLR and molecular torus.}
\label{fig:Pol_LC_torus}
\end{figure}

In order to create a more physical model of X-ray eclipses, we decided to also account for a more realistic geometrical scheme. In addition to the hot corona and the accretion disc, we consider the contribution of an additional equatorial molecular torus and a biconical NLR to the observed polarized light that is scattered along the observer's line of sight, using the STOKES code\footnote{\url{http://stokes-program.info/}}. STOKES is a Monte Carlo code developed by \cite{Goosmann07}, \cite{Marin12, Marin15b} and \cite{Rojas18} that allows us to simulate the radiative transfer of photons in a wide three-dimensional environment. The code accounts for all the physics of scattering, absorption and re-emission from the near-infrared to the hard X-ray band (with the exception of magnetically-driven mechanisms such as synchrotron emission or dichroism that are not relevant here). Similarly to the work carried out by \cite{Marin18, Marin18b}, it is possible to use the output from the KY code as input for STOKES. The photons can then propagate in a parsec-scale environment, allowing for a realistic radiative coupling between the central engine and the parsec-scale AGN structure. In this case, we followed the work of \cite{Marin18, Marin18b} for Seyferts and included a circumnuclear torus with a half-opening angle of 30\degr\ from the equatorial plane and a hydrogen column density of $10^{24}\,\rm cm^{-2}$. The inner radius of the torus is set to a distance of 0.01~pc according to reverberation mapping data \citep{Suganuma06, Vazquez15}, and the outer one to 5~pc. The NLR is bi-conical, optically thin, and have a hydrogen column density of $10^{21}\,\rm cm^{-2}$. They represent the typical NLR detected in many AGN and extend up to 60 parsecs before mixing with the galactic environment.

We present in Fig.~\ref{fig:Pol_LC_torus} the temporal variations of the polarization parameters that include the extra contributions of the torus and the NLR, in the 4-8 keV and 20-50 keV bands. We also plot the parameters for the same configurations considering only the cloud-corona-disc system, for comparison. We considered only the case of unpolarized primary at $h=2.5\,\rm r_g$. First, we note that, for all cases, the inclusion of the contribution from the torus and NLR results in an enhancement in the relative observed intensity during the eclipse. In fact, the contribution of these components to the scattered light in the line of sight is constant for the whole event and not affected by the eclipse itself. As expected, in the unobscured cases, the polarization degree increases when the contributions from the torus and NLR are included, especially for higher energies. The inclusion of torus and NLR result in a small increase in $P(E)$ during the unobscured phase. This is in agreement with the results in \cite{Marin18,Marin18b} where the authors investigate the effect of the molecular torus and NLR on the polarization signal for type-2 and type-1 Seyferts, respectively. However, this contribution, being stable and non-negligible in terms of intensity, reduces the amplitude of variability, once the primary is obscured. In fact, adding this contribution significantly dilutes the polarization signal leading to lower polarization degrees with respect to the cases when we do not account for these components, thus lower variability amplitude. During the eclipse, $P(t)$ becomes $\sim 1.3/2$\% (7/8\%) for the Schwarzchild/Kerr BHs in the 4--8~keV (20--50~keV) energy bands for an inclination of 30\degr. While for an inclination of 60\degr, $P(t)$ becomes $\sim 3/5\%$ for the Schwarzchild/Kerr BHs in the 4--8~keV band, while it is $\sim 16\%$ in the 20--50~keV for both spins. We note, however, that qualitatively the same variability pattern can be observed with and without the inclusion of the contribution from the parsec-scale material. The same is seen when comparing the variability patterns of $\Psi(t)$, with the only difference that the inclusion of the torus and NLR tends to reduce the values of the polarization position angles, leading to a more parallel polarization.

\section{Discussion}
\label{sec:discussion}

We showed that X-ray eclipsing events can be used in order to probe the emission from the innermost regions of AGN. This had been investigated earlier by \cite{Risaliti11}, \cite{Marin15} and \cite{Sanfrutos16}. However, we present in this work a new analysis taking into account full relativistic effects and more complex and physical configurations compared to the ones assumed in the aforementioned works. We note that our results depend on several uncertainties, mainly caused by our poor knowledge of the exact geometry of the system. 

\begin{enumerate}
	\item An extended corona would be a more realistic case, and may lead to different spectral and 
	polarimetric signatures with respect to the lamp-post geometry that is assumed in this work. 
	This would mainly dilute the sudden changes caused by the high-compactness 
	of the lamp-post. One would expect to observe more gradual patterns 
	during the obscuration of the primary \citep[see e.g.][]{Sanfrutos16}. 
	However, we note that models with such a geometry are still not available for X-ray 
	polarization.
	
	\item The primary emission, assumed to be a power-law spectrum, is also expected 
	to vary in flux and/or shape during the observation, which we do not take into consideration in 
	this work.
	
	\item The structure and geometry of the obscuring material may be more complex having a 
	gradient of column density along the line of sight and a non-spherical shape 
	\citep{Maiolino10}. We also note that the column density of the obscuring cloud may 
	play a role in modifying mainly 
	the spectral signatures. Obscuration caused by clouds with lower column densities, 
	on the order of $\sim 10^{23}~\rm cm^{-2}$ \citep[e.g.][]{Risaliti09, Maiolino10}, 
	will mostly affect the soft X-rays.
	
\end{enumerate}

Our spectral simulations are promising for the future high-resolution X-ray micro-calorimeters on board of ATHENA \citep{Athena} and, possibly, earlier X-ray missions such as Arcus \citep{Arcus} and XARM. The spectra that will be provided by the micro-calorimeters would be of great help in order to track the variability of fine spectral structures (especially the Fe line or the soft bands), during the eclipsing event. This would give us strong hints on the local emissivity of the disc, thus on the geometry of the corona. The detectability of these events and their implications will be investigated in future work. However, we note that it is not yet plausible to detect the effects of obscuration on the polarimetric signals in AGN,  on short timescales. In fact, \cite{Marin12MCG} showed that more than $\sim 1$~Ms will be needed for a XIPE-like S-class mission assuming a non variable source with a flux of 3~mCrab in the 2--10~keV band. This is much larger than the timescale of the obscuration event in which the flux drops drastically, making the measurements even harder to be achieved. Instead, it would be possible with the up-coming X-ray polarimeters such as eXTP \citep{Zhang16} to catch changing-look (on larger timescales) sources in two different obscured/unobscured states. Sources with high mass and/or large obscuring clouds would be optimal for such analysis, as the duration of the eclipse would be longer in these cases. Combining the spectral and polarimetric capabilities of future missions would allow us not only to probe the innermost regions of the AGN, but it may give us also strong hints about the structure of the parsec-scale material. Our results clearly show that taking into consideration the contribution from this scattering material would alter the polarization signal in terms of polarization degree and position angle, as well as variability. These effects strongly depend on the structure and the geometry of the parsec-scale material.

\section{Conclusions}
\label{sec:conclusions}

We have shown in this paper how X-ray eclipses in AGN can affect the observed spectral and polarimetric signals, being then a powerful means to probe the relativistic effects that dominate the innermost regions of these sources. Our main conclusions are the following.

\begin{enumerate}

	\item The observed X-ray spectra show asymmetries during the different phases of the eclipse
	as the cloud is shading various parts of the accretion disc. These effects depend strongly on 
	the location of the primary source and the size of the obscuring cloud. We also showed that 
	for large heights it becomes harder to determine the BH spin.
	
	\item Considering the corona-disc system only, an asymmetric enhancement in the polarization 
	signal is expected to occur as the cloud is shading the innermost regions of the system. This 
	enhancement is highly dependent on the inclination of the system: the higher the inclination, 
	the larger the polarization degree. In addition, the variability patterns of the polarization 
	degree and the polarization position angle depend strongly on the location of the primary. 
	We also showed that the effect of the spin is less prominent. However, the temporal evolution 
	of the polarization position angle is highly affected by the intrinsic polarization of the 
	primary source.
	
	\item It is crucial to consider a full geometrical configuration of the AGN. 
	Accounting for the polarization signal from parsec-scale AGN components, the expected 
	total polarization from the whole system is strongly altered. The inclusion of the 
	constant scattered light by the torus and NLR in the line of sight tends to increase 
	the degree of polarization when the innermost regions are unobscured. However, once 
	the corona is obscured, the contribution from the torus and NLR, which is not 
	affected by the eclipsing event, tends to smooth out the variability in polarization. This 
	reduces remarkably the degree of polarization, and leads to a more parallel polarization 
	($\Psi$ closer to 0). {We note that, despite the fact that in this case 
	the variability of the polarization signal is lower, 
	its absolute value can provide unique information on the geometry 
	and the properties of the parsec-scale scattering medium.}

\end{enumerate}

In the future we will investigate in more detail the observational prospects of spectral variability caused by obscuration events that will be detected by the next generation of X-ray observatories. We also aim at considering more realistic and complex configurations of the AGN. Moreover, it is of particular interest to consider the impact of an extended corona on the observed spectral and polarimetric signatures, such as recently demonstrated by \cite{Chauvin18}. Finally, we will also look at the effects of varying the characteristics of the accretion disc.

\section*{Acknowledgements}

We are grateful to Dr. Giovanni Miniutti for providing valuable comments and suggestions that improved the manuscript. ESK acknowledges the support by the ERASMUS+ Programme - Student Mobility for Traineeship-Project KTEU-ET (Key to Europe Erasmus Traineeship). ESK thanks the Astronomical Insitute of the Czech Academy of Sciences for hospitality during the early phase of this work. FM is grateful to the Centre national d'\'{e}tudes spatiales (CNES) which provided research funding though the post-doctoral grant ``Probing the geometry and physics of active galactic nuclei with ultraviolet and X-ray polarized radiative transfer''. MD thanks MEYS of Czech Republic for the support through the 18-00533S project. EN received funding from the European Union's Horizon 2020 research and innovation programme under the Marie Sk\l odowska-Curie grant agreement no. 664931. MS acknowledges support from the DGAPA-UNAM postdoctoral fellowships program. The figures were generated using {\tt matplotlib} \citep{Hunter07}, a {\tt PYTHON} library for publication of quality graphics. 

\bibliographystyle{mnras}
\bibliography{ek-obscuration-ref} 



\bsp	
\label{lastpage}
\end{document}